# Adaptive Inference through Bayesian and Inverse Bayesian Inference with Symmetry Bias in Nonstationary Environments

Shuji Shinohara[a,*], Daiki Morita[a], Hayato Hirai[a], Ryosuke Kuribayashi[a], Nobuhito Manome[b,c], Toru Moriyama[d], Yoshihiro Nakajima[e], Yukio-Pegio Gunji[f], and Ung-il Chung[b]

[a] School of Science and Engineering, Tokyo Denki University, Saitama, Japan

[b] Department of Bioengineering, Graduate School of Engineering, The University of Tokyo, Tokyo, Japan

[c] Department of Research and Development, SoftBank Robotics Group Corp., Tokyo, Japan

[d] Faculty of Textile Science, Shinshu University, Ueda, Japan

[e] Graduate School of Economics, Osaka Metropolitan University, Osaka, Japan

[f] Department of Intermedia Art and Science, School of Fundamental Science and Technology, Waseda University, Tokyo, Japan

*** Corresponding author**

**E-mail:** s.shinohara@mail.dendai.ac.jp

**Postal address:** School of Science and Engineering, Tokyo Denki University, Ishizaka, Hatoyama-machi, Hiki-gun, Saitama 350-0394, Japan

**Abstract**

This study proposes the novel Bayesian and inverse Bayesian (BIB) inference framework that incorporates symmetry bias into the Bayesian updating process to perform both conventional and inverse Bayesian updates concurrently. Conventional Bayesian inference is constrained by a fundamental trade-off between adaptability to abrupt environmental changes and accuracy during stable periods. The BIB framework addresses this limitation by dynamically modulating the learning rate via inverse Bayesian updates, thereby enhancing adaptive flexibility. The BIB model was evaluated in a sequential estimation task involving observations drawn from a Gaussian distribution with a stochastically time-varying mean, where it exhibited spontaneous bursts in the learning rate during environmental transitions, transiently entering high-sensitivity states that facilitated rapid adaptation. This burst–relaxation dynamic serves as a mechanism for balancing adaptability and accuracy. Furthermore, avalanche analysis, detrended fluctuation analysis, and power spectral analysis revealed that the BIB system likely operates near a critical state—a property not observed in standard Bayesian inference. This suggests that the BIB model uniquely achieves a coexistence of computational efficiency and critical dynamics, resolving the adaptability–accuracy trade-off while maintaining scale-free behavior. These findings offer a new computational perspective on scale-free dynamics in natural systems and provide valuable insights for the design of adaptive inference systems in nonstationary environments.

## 1 Introduction

Prediction and action selection based on external information are fundamental components of human cognition. In recent years, such predictive behaviors have increasingly been interpreted through the frameworks of the free energy principle, active inference, and prediction error minimization [1–3]. The free energy principle in particular is formally grounded in Bayesian inference [1, 4], and the role of Bayesian mechanisms in brain function has attracted considerable interest in contemporary neuroscience [5, 6].

Bayesian inference provides a principled computational framework for estimating hidden causes from sensory data by recursively updating posterior beliefs on the basis of prior knowledge and incoming observations. However, human reasoning often deviates from this normative framework and results in systematic cognitive biases. For example, when presented with a unidirectional conditional statement ("if p, then q"), individuals frequently infer its converse ("if q, then p") or its contrapositive ("if not p, then not q") as well. These tendencies, known as symmetry bias and mutual exclusivity bias, are considered fundamental characteristics of intuitive human reasoning [7–10].

In the domain of causal induction, it has been shown that the perceived strength of a causal relationship between a cause c and an effect e depends not only on the predictive probability, $P(e|c)$ but also on the diagnostic probability, $P(c|e)$. The geometric mean of these two probabilities has been proposed as a psychologically plausible measure of causal strength [11], underscoring the importance of symmetry in human causal reasoning.

Although the accumulation of observational data typically improves inference accuracy, this assumption primarily holds in stationary environments. In nonstationary contexts where causal relationships evolve over time, inference must dynamically adapt to recent changes. This creates a fundamental trade-off between stability (accuracy) and flexibility (adaptability). To address this issue, Shinohara et al. [12]

proposed a Bayesian inference model that incorporates symmetry bias, along with dynamic mechanisms such as memory decay and flexible likelihood updates, enabling adaptive inference in changing environments.

Conventional Bayesian updating sharpens belief distributions by reducing the posterior variance. Gunji et al. [13] proposed inverse Bayesian updating, a complementary mechanism that broadens the hypothesis space by reversing the direction of inference from data to hypotheses. The integration of both Bayesian and inverse Bayesian processes enables flexible belief revision and has been applied to model complex behaviors, such as collective animal movement and human decision-making [14–16]. However, previous studies have predominantly assumed a discrete hypothesis space, which limits their applicability to problems requiring continuous estimation. Furthermore, the relationship between inverse Bayesian inference and symmetry bias remains poorly understood. To address these limitations, we propose a novel theoretical model of Bayesian inference that integrates symmetry bias within a continuous Gaussian prior framework. The model includes a dynamic mechanism that expands the likelihood variance through symmetry bias, enabling both sharpening of beliefs (via Bayesian updating) and broadening of the hypothesis space (via inverse Bayesian updating). When applied to a Gaussian mean estimation task in a stochastically changing environment, the model effectively mitigates the conventional trade-off between inference accuracy and adaptability.

In recent years, increasing empirical evidence has suggested that the brain may operate near a critical state, a regime located at the boundary between order and disorder [17]. Criticality has been associated with several functional advantages, including optimal information transmission, heightened sensitivity to external inputs, and enhanced computational capacity [18, 19]. A hallmark of neural criticality is the presence of neuronal avalanches, which are bursts of activity whose size $S$ and duration $T$ follow power-law distributions. These distributions adhere to scaling laws of the form $P(S) \sim S^{-\tau}$, $P(T) \sim T^{-\tau_t}$ and the average size–duration relationship $\langle S \rangle_T \sim T^{\gamma}$, where the critical exponents satisfy the scaling relation $\gamma = \frac{\tau_t - 1}{\tau - 1}$, which is consistent with a critical branching process [20, 21].

Such avalanche dynamics have been observed in resting-state brain activity across various modalities (EEG, MEG, and fMRI) in both humans and animals [22], and they persist during active cognitive tasks such as decision-making and inference [23, 24]. Moreover, behavioral performance has been shown to correlate with markers of criticality, such as avalanche exponents and long-range temporal correlations (LRTCs), suggesting that critical dynamics may support cognitive flexibility and adaptive behavior.

In this study, we performed avalanche analysis, detrended fluctuation analysis (DFA) and power spectral analysis (PSA) on time series data generated via the inference dynamics of our proposed model. Specifically, we examined the distributions of the avalanche size $S$ and duration $T$, along with their scaling relationships, to determine whether the model exhibits signatures of criticality. In addition, DFA and PSA were used to assess the presence or absence of LRTC. This approach was adopted to elucidate the potential role of critical dynamics as a foundational principle underlying adaptive inference and contributes to a deeper understanding of scale-free behavior in cognitive systems.

## 2 Materials and methods

### 2.1 Bayesian and inverse Bayesian (BIB) inference with symmetry bias

We introduce a symmetry bias as follows based on the study by Shinohara et al. [9, 12].

$$P_{t+1}(h \mid d) := \frac{1}{Z_h} P_t(h \mid d)^{1-\beta} P_t(d \mid h)^{\beta} \tag{1}$$

$$P_{t+1}(d \mid h) := \frac{1}{Z_d} P_t(d \mid h)^{1-\beta} P_t(h \mid d)^{\beta} \tag{2}$$

where $h$ denotes a hypothesis and $d$ represents data. $Z_h$ and $Z_d$ denote normalization constants, and $P_t(d \mid h)$ and $P_t(h \mid d)$ represent the likelihood and posterior probability at time $t$, respectively. The parameter $0 \le \beta \le 1$ represents the strength of the symmetry bias. Shinohara et al. [9, 12] defined $P_{t+1}(h \mid d)$ as a weighted generalized mean of $P_t(d \mid h)$ and $P_t(h \mid d)$; however, in this study, we employ a weighted geometric mean as a special case for simplicity. When $\beta = 0$, Eqs. (1) and (2) become an identity each, and nothing changes. When $\beta = 0.5$, Eqs. (1) and (2) become $P_{t+1}(h \mid d) \propto \sqrt{P_t(d \mid h) P_t(h \mid d)}$, $P_{t+1}(d \mid h) \propto \sqrt{P_t(d \mid h) P_t(h \mid d)}$, and symmetry $P_{t+1}(d \mid h) \propto P_{t+1}(h \mid d)$ holds. Eqs. (1) and (2) can be transformed as follows using Bayes' theorem $P_t(h \mid d) = \dfrac{P_t(h) P_t(d \mid h)}{P_t(d)}$.

$$P_{t+1}(h \mid d) := \frac{1}{Z_h} \left( \frac{P_t(h)}{P_t(d)} \right)^{1-\beta} P_t(d \mid h) \tag{3}$$

$$P_{t+1}(d \mid h) := \frac{1}{Z_d} \left( \frac{P_t(h)}{P_t(d)} \right)^{\beta} P_t(d \mid h) \tag{4}$$

Using Bayesian update, we rewrite Eq. (3) as follows:

$$P_{t+1}(h) := \frac{1}{Z_h} \left( \frac{P_t(h)}{P_t(d_t)} \right)^{1-\beta} P_t(d_t \mid h) \tag{5}$$

where $d_t$ represents the observed data at time *t*.

When $\beta = 0$, Eq. (5) reduces to standard Bayesian update, and Eq. (4) becomes $P_{t+1}(d_t \mid h) = P_t(d_t \mid h)$; the likelihood distribution does not change. In Eq. (5), $P_t(d_t)$ is common across all the hypotheses and therefore can be considered a constant. Therefore, $P_{t+1}(h) \propto P_0(h)^{(1-\beta)^t} \cdots P_{t-i}(d_{t-i} \mid h)^{(1-\beta)^i} \cdots P_t(d_t \mid h) = P_0(h)^{(1-\beta)^t} \prod_{i=0}^{t-1} P_{t-i}(d_{t-i} \mid h)^{(1-\beta)^i}$ holds owing to the recursiveness of $P(h)$ [12]. This formulation implies that likelihoods corresponding to older observations are progressively discounted when $\beta > 0$, exerting a decreasing influence on the current prior distribution. Hence, β can be interpreted as a discount factor or forgetting rate.

Consider the use of Bayesian inference for estimating the mean of a Gaussian distribution that generates the observed data. Let the hypothesis *h* represent the mean μ of the Gaussian distribution. Define the prior distribution of μ as

$$P_t(\mu) = \frac{1}{\sqrt{2\pi\Phi_t}} \exp\left[ -\frac{1}{2\Phi_t} (\mu - \theta_t)^2 \right] \tag{6}$$

where $\Phi_t$ and $\theta_t$ denote the variance and mean of the prior distribution, respectively. The term $\mu = \theta_t$ corresponds to the most confident hypothesis at time *t*, that is, the current estimate. The likelihood is also modeled as a Gaussian distribution:

$$P_t(d \mid \mu) = \frac{1}{\sqrt{2\pi\Sigma_t}} \exp\left[ -\frac{1}{2\Sigma_t} (d - \mu)^2 \right] \tag{7}$$

where $\Sigma_t$ denotes the likelihood variance.

Substituting $P_t(h)$ and $P_t(d_t \mid h)$ in Eq. (5) with $P_t(\mu)$ and $P_t(d_t \mid \mu)$ in Eqs. (6) and (7), respectively, yields

$$\begin{aligned} P_{t+1}(\mu) &\propto \left[ P_t(\mu) \right]^{1-\beta} P_t(d_t \mid \mu) \\ &\propto \left[ \exp\left[ -\frac{1}{2\Phi_t} (\mu - \theta_t)^2 \right] \right]^{1-\beta} \exp\left[ -\frac{1}{2\Sigma_t} (d_t - \mu)^2 \right]. \\ &= \exp\left[ -\frac{1}{2} \left[ \frac{(1-\beta)(\mu - \theta_t)^2}{\Phi_t} + \frac{(d_t - \mu)^2}{\Sigma_t} \right] \right] \end{aligned} \tag{8}$$

The prior distribution can be updated by rearranging Eq. (8) as follows:

$$\begin{aligned} P_{t+1}(\mu) &= \frac{1}{\sqrt{2\pi\Phi_{t+1}}} \exp\left[ -\frac{1}{2\Phi_{t+1}} \left[ \mu - \theta_{t+1} \right]^2 \right] \\ \Phi_{t+1} &= \left( \frac{(1-\beta)}{\Phi_t} + \frac{1}{\Sigma_t} \right)^{-1} = \frac{\Phi_t}{\Phi_t + (1-\beta)\Sigma_t} \Sigma_t \ . \\ \theta_{t+1} &= \frac{(1-\beta)\Sigma_t}{\Phi_t + (1-\beta)\Sigma_t} \theta_t + \frac{\Phi_t}{\Phi_t + (1-\beta)\Sigma_t} d_t \end{aligned} \tag{9}$$

Let $\alpha_{t+1} = \frac{\Phi_t}{\Phi_t + (1-\beta)\Sigma_t}$ denote the learning rate. $\Phi_{t+1}$ and $\theta_{t+1}$ can be rewritten as follows using the learning rate.

$$\Phi_{t+1} = \alpha_{t+1} \Sigma_t \tag{10}$$

$$\theta_{t+1} = \left(1-\alpha_{t+1}\right)\theta_t + \alpha_{t+1} d_t \tag{11}$$

These results confirm that Eq. (5) describes Bayesian updating with a discount factor. However, the interpretation of Eq. (4) remains to be addressed. Substituting $P_t\left(\mu\right)$ and $P_t\left(d \mid \mu\right)$ with $P_t\left(h\right)$ and $P_t\left(d \mid h\right)$, Eq. (4) becomes

$$\begin{aligned}
& P_{t+1}\left(d \mid \mu\right) \propto \left[\frac{P_t\left(\mu\right)}{P_t\left(d\right)}\right]^{\beta} P_t\left(d \mid \mu\right) \\
& \propto \left[\exp\left[-\frac{1}{2\Phi_t}\left(\mu-\theta_t\right)^2\right]\right]^{\beta} \exp\left[-\frac{1}{2\left(\Sigma_t+\Phi_t\right)}\left(d-\theta_t\right)^2\right]^{-\beta} \exp\left[-\frac{1}{2\Sigma_t}\left(d-\mu\right)^2\right] \\
& = \exp\left[-\frac{1}{2}\left[\frac{\beta}{\Phi_t}\left(\mu-\theta_t\right)^2 - \frac{\beta}{\Sigma_t+\Phi_t}\left(d-\theta_t\right)^2 + \frac{\left(d-\mu\right)^2}{\Sigma_t}\right]\right] \\
& = \exp\left[-\frac{1}{2}\left[\left(\frac{1}{\Sigma_t} - \frac{\beta}{\Sigma_t+\Phi_t}\right)d^2 - 2\left(\frac{\mu}{\Sigma_t} - \frac{\beta\theta_t}{\Sigma_t+\Phi_t}\right)d - \frac{\beta\theta_t^2}{\Sigma_t+\Phi_t} + \frac{\mu^2}{\Sigma_t} + \frac{\beta}{\Phi_t}\left(\mu-\theta_t\right)^2\right]\right] \\
& \propto \exp\left[-\frac{1}{2}\left[\left(\frac{1}{\Sigma_t} - \frac{\beta}{\Sigma_t+\Phi_t}\right)d^2 - 2\left(\frac{\mu}{\Sigma_t} - \frac{\beta\theta_t}{\Sigma_t+\Phi_t}\right)d\right]\right] \\
& \propto \exp\left[-\frac{1}{2}\left(\frac{1}{\Sigma_t} - \frac{\beta}{\Sigma_t+\Phi_t}\right)\left[d - \left(\frac{1}{\Sigma_t} - \frac{\beta}{\Sigma_t+\Phi_t}\right)^{-1}\left(\frac{\mu}{\Sigma_t} - \frac{\beta\theta_t}{\Sigma_t+\Phi_t}\right)\right]^2\right]
\end{aligned} \tag{12}$$

where $P_t\left(d\right) = \int P_t\left(\mu\right) P_t\left(d \mid \mu\right) d\mu = \frac{1}{\sqrt{2\pi\left(\Sigma_t+\Phi_t\right)}} \exp\left[-\frac{1}{2\left(\Sigma_t+\Phi_t\right)}\left(d-\theta_t\right)^2\right]$ was used.

The likelihood can be written as follows by rearranging Eq. (12):

$$\begin{aligned}
& P_{t+1}\left(d \mid \mu'\right) = \frac{1}{\sqrt{2\pi\Sigma_{t+1}}} \exp\left[-\frac{1}{2\Sigma_{t+1}}\left(d-\mu'\right)^2\right] \\
& \mu' = \Sigma_{t+1}\left(\frac{\mu}{\Sigma_t} - \frac{\beta\theta_t}{\Sigma_t+\Phi_t}\right) \\
& \Sigma_{t+1} = \left(\frac{1}{\Sigma_t} - \frac{\beta}{\Sigma_t+\Phi_t}\right)^{-1} = \frac{\Phi_t+\Sigma_t}{\Phi_t+\left(1-\beta\right)\Sigma_t}\Sigma_t
\end{aligned} \quad . \tag{13}$$

Given that μ is a continuous variable, replacing $\mu'$ with μ in Eq. (13) yields

$$
\begin{aligned}
P_{t+1}(d \mid \mu) &= \frac{1}{\sqrt{2\pi\Sigma_{t+1}}}\exp\left[-\frac{1}{2\Sigma_{t+1}}(d-\mu)^2\right] \\
\Sigma_{t+1} &= \frac{\Phi_t+\Sigma_t}{\Phi_t+(1-\beta)\Sigma_t}\Sigma_t
\end{aligned}
\quad (14)
$$

If $\beta > 0$, then $\frac{\Phi_t+\Sigma_t}{\Phi_t+(1-\beta)\Sigma_t} > 1$, resulting in an increasing likelihood variance Σ. Hence, Eq. (14) represents a variance-expanding update. Hereafter, this likelihood update process is referred to as an inverse Bayesian update.

Gunji et al. [13] defined inverse Bayesian updating as $P_{t+1}(d \mid \mu) := P_t(d)$. As noted earlier, $P_t(d) = \int P_t(\mu)P_t(d \mid \mu)d\mu = \frac{1}{\sqrt{2\pi(\Sigma_t+\Phi_t)}}\exp\left[-\frac{1}{2(\Sigma_t+\Phi_t)}(d-\theta_t)^2\right]$ holds, where $P_t(d)$ represents the weighted average of the individual likelihoods $P_t(d \mid \mu)$. Inverse Bayesian updating involves substituting individual likelihoods with the global average. The variance of $P_t(d \mid \mu)$ is $\Sigma_t$, whereas that of $P_t(d)$ is $\Sigma_t + \Phi_t$. Therefore, inverse Bayesian updating can be interpreted as a process that expands the likelihood variance from $\Sigma_t$ to $\Sigma_t + \Phi_t$.

Compared with Eq. (14), both processes increase the variance of the likelihood. Gunji's definition of inverse Bayes updating corresponds to the case where $\beta = \frac{\Phi_t}{\Sigma_t}$ is used in Eq. (14). Hereafter, the method involving only Bayesian updating with a discount rate, governed by Eqs. (10) and (11), is referred to as Bayesian inference. However, the proposed model, which combines Bayesian and inverse Bayesian updates by incorporating Eq. (14), is referred to as BIB inference.

Under Bayesian inference, Eq. (14) is not used to update the likelihood; hence, the likelihood remains fixed at $\Sigma_t = \Sigma_0$. In this case, $\Phi_{t+1} = \alpha_{t+1}\Sigma_0 \propto \alpha_{t+1}$ holds from Eq. (10), and the variances of the prior distribution are proportional to the learning rate. Specifically, the prior variance is given by $\Phi_{t+1} = \alpha_{t+1}\Sigma_0 = \frac{\Phi_t}{\Phi_t + (1-\beta)\Sigma_0}\Sigma_0$ , and the learning rate is $\alpha_{t+1} = \frac{\Phi_t}{\Phi_t + (1-\beta)\Sigma_0}$ , implying that $\lim_{t\to\infty}\Phi_t = \beta\Sigma_0$ ; hence, $\lim_{t\to\infty}\alpha_t = \beta$ . Specifically, the learning and discount rates reach equilibrium over time.

When the discount rate β is 0, both the prior variance and the learning rate asymptotically converge to zero. This behavior reflects increasing confidence in the estimate $\theta_t$ as more data are accumulated, ultimately leading to the exclusion of new information from subsequent observations. However, when a nonzero discount rate is applied ( $\beta > 0$ ), the variance of the prior distribution and the learning rate do not become zero because the influence of past observations is limited by the forgetting mechanism. Consequently, new data continue to inform the estimate at a constant rate determined by the discount factor β.

The likelihood variance $\Sigma_t$ increases with each update in BIB inference. With respect to the prior variance, $\Phi_{t+1} = \alpha_{t+1}\Sigma_t = \frac{\Phi_t}{\Phi_t + (1-\beta)\Sigma_t}\Sigma_t = \frac{1}{\Phi_t/\Sigma_t - \beta + 1}\Phi_t$ holds. Hence, after the point at which $\Sigma_t$ increases and $\frac{\Phi_t}{\Sigma_t} < \beta$ holds, $\Phi_t$ also increases and diverges over time. Consequently, the prior distribution becomes uniform over all values of μ, effectively eliminating distinctions among hypotheses. Hence, BIB inference performs two mutually opposing operations: Bayesian updating progressively narrows the set of plausible hypotheses on the basis of observed data, whereas inverse Bayesian updating disrupts the foundational premise of Bayesian inference, namely, the identification of a single, most plausible hypothesis.

Here, we introduce a set of discrete reference points $\mu^0, \mu^1, \cdots, \mu^i, \cdots, \mu^N$ to detect changes in the data-generating distribution. We define the hypothesis with the highest confidence at time *t* as $\mu_t^{\max} = \arg\max\left(P_t\left(\mu^i\right)\right)$. We consider the generating distribution to change in the case of $\mu_{t+1}^{\max} \neq \mu_t^{\max}$ and reset the variance to $\Sigma_t = \Sigma_0$ .

In both cases, the hypothesis with the highest prior confidence may change. The first occurs when the observed data significantly deviate from the previously estimated values. According to Eq. (11), a large discrepancy between the estimate $\theta_t$ and the observed data $d_t$, combined with a high learning rate $\alpha_{t+1}$, results in a substantial update to $\theta_t$. In such cases, the prior distribution of $P_{t+1}(\mu)$ changes substantially, potentially altering the most confident hypothesis. This change can be interpreted as anomaly detection in the incoming data. The second case is when the variances of the likelihood and the prior diverge because of the inverse Bayes update. In this case, the confidence levels across hypotheses become indistinguishable owing to $P(\mu^0) = P(\mu^1) = \cdots = P(\mu^i) = \cdots = P(\mu^N) = 0$, and the most confident hypothesis is selected randomly.

Specifically, when the likelihood variance diverges, the likelihood function no longer satisfies the definition of a proper probability distribution. This condition does not imply that all hypotheses become equally plausible but rather that all hypotheses effectively vanish. The inability to determine a single hypothesis with maximal confidence stems not from multiple hypotheses having the same confidence level but from the fact that the very problem of selecting an optimal hypothesis becomes ill-defined.

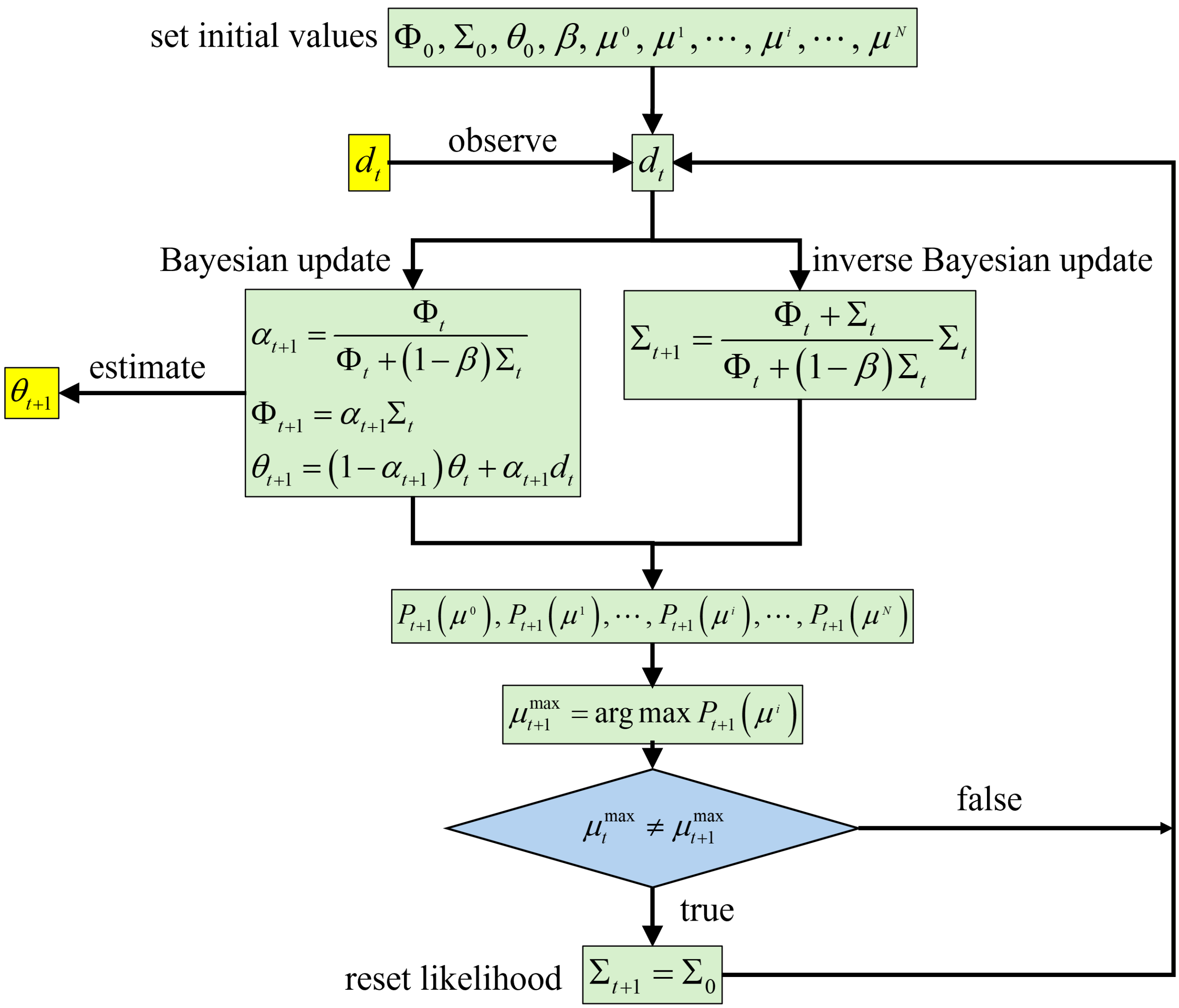

**Fig. 1** Computational procedures for Bayesian inference and BIB inference. Inverse Bayesian update was not performed in Bayesian inference.

Fig. 1 illustrates the procedures for Bayesian and BIB inferences. In Bayesian inference without inverse Bayesian updating, resets occur frequently owing to the first scenario. However, the inference process remains unaffected by resets because the likelihood variance remains fixed. However, a burst in the learning rate occurs in the BIB inference when the variance $\Sigma_t$ of the expanded likelihood is reset to $\Sigma_0$.

### 2.2 Simulation setting

This study simulated the task of estimating the mean of a normal distribution, generating observed data in a nonstationary environment.

The data-generating distribution was assumed to follow the normal distribution defined as

$$P_t(d) = \frac{1}{\sqrt{2\pi\Omega}} \exp\left[ -\frac{1}{2\Omega}(d - \eta_t)^2 \right] \tag{15}$$

The variance was fixed at $\Omega = 0.09$. In terms of the mean, we performed the simulations using two settings. First, we varied the mean $\eta_t$ irregularly to check the tracking performance of the system. Specifically, a random number $rnd$ was sampled from the uniform distribution $U(0,1)$ at each time step. If $rnd < 0.001$, the mean is changed. In such cases, a random number $rnd\eta$ was sampled from the uniform distribution $U(-2.5, 2.5)$, which was then adopted as the new mean.

$$\eta_{t+1} = \begin{cases} rnd\eta \sim U(-2.5, 2.5) & if\ rnd < 0.001, \quad rnd \sim U(0,1) \\ \eta_t & otherwise \end{cases} \tag{16}$$

This is referred to as Simulation 1. The second setting was fixed at $\eta_t = 0$ to examine the criticality of the system. This is referred to as Simulation 2.

For the discrete hypothesis reference points $\mu^0, \mu^1, \cdots, \mu^i, \cdots, \mu^N$, $N = 50$, and the interval between the points is $\Delta\mu = 0.1$, fixed at $\mu^i = -2.5 + i\Delta\mu$.

Data from the initial 10,000 time steps were excluded from the analysis to eliminate the influence of transient behavior. Only data from time step 10,001 onward were used.

The simulation was implemented in C++ via the MinGW11.2.0 64-bit compiler [25]. The Qt library (Qt version Qt 6.4.0 MinGW 64-bit) was also used for the development [26]. The maximum representative real number in this simulation environment was 1.79769e+308. Values exceeding this limit were considered infinity, and their reciprocals were recognized as 0.

# 3 Results

### 3.1 Tracking performance analyses

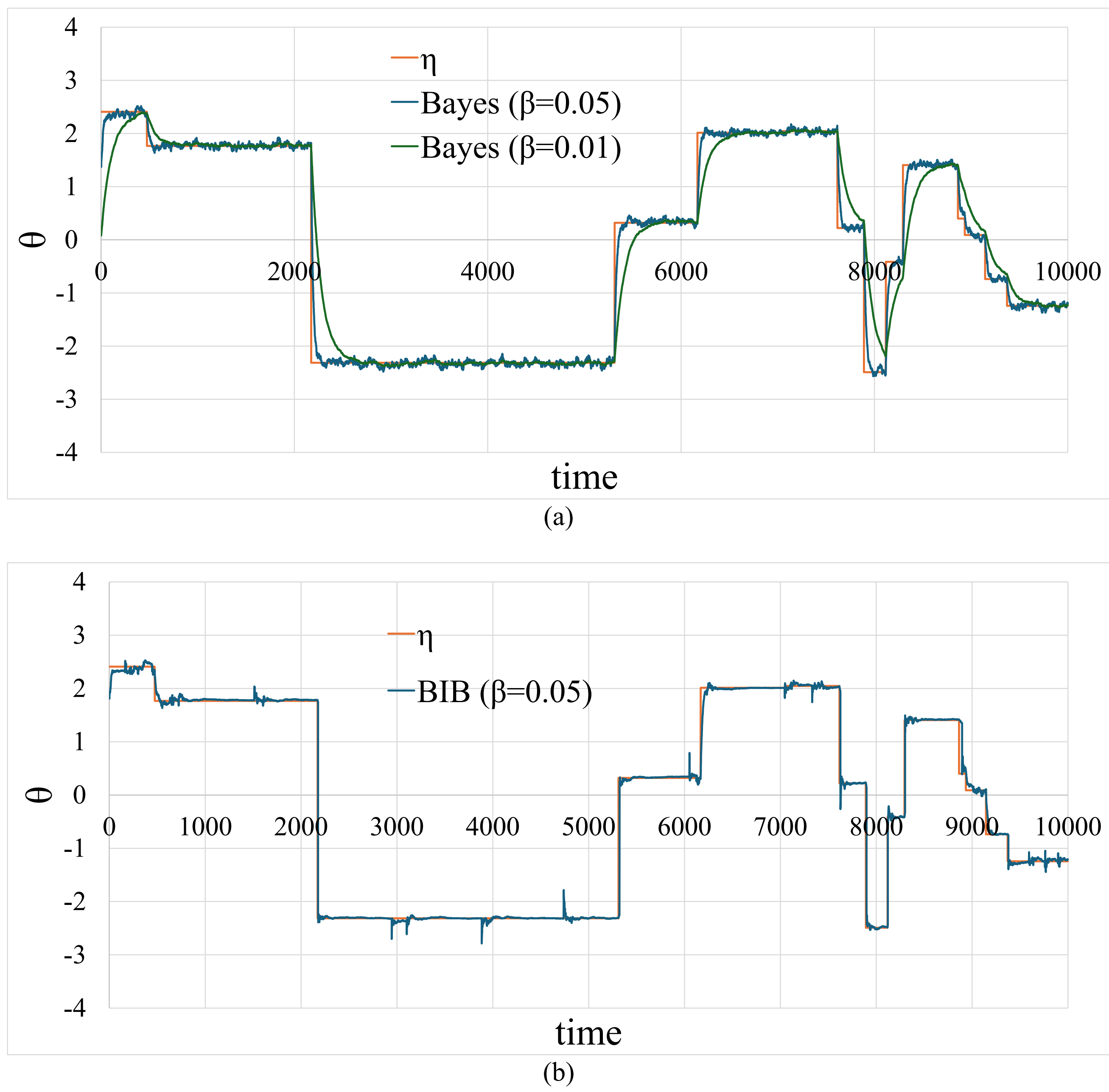

**Fig. 2** Representative time series of the mean value η of the data-generating distribution and corresponding estimates θ in simulation 1. (a) Estimates obtained via Bayesian inference ( $\beta = 0.05$ , $\beta = 0.01$ ). (b) Estimates obtained using BIB inference ( $\beta = 0.05$ ).

Figs. 2(a) and 2(b) depict examples of the estimated mean values of the data-generating distributions obtained via Bayesian inference and BIB inference, respectively. Each figure includes the true mean value η of the generating distribution for reference. The Bayesian estimation (Fig. 2(a)) demonstrates that increasing the discount rate β enhances responsiveness to abrupt changes but reduces accuracy during periods of stability. However, BIB estimation occasionally results in bursts in the estimated values while achieving both high tracking performance and accuracy.

The interval $\left[t', t'+n\right]$ from a specific time $t'$, when the mean of the data-generating distribution changes, to the time immediately preceding the next change was divided into two equal halves to quantify adaptability to change and estimation accuracy. The root mean square error (RMSE) between the estimated and true values was calculated separately for each half. Specifically, the RMSEs were denoted as $RMSE_f = \sqrt{\frac{2}{n}\sum_{i=t'}^{t'+n/2-1}\left(\theta_i - \eta_i\right)^2}$ and $RMSE_s = \sqrt{\frac{2}{n}\sum_{i=t'+n/2}^{t'+n}\left(\theta_i - \eta_i\right)^2}$ for the first and second halves, respectively. These RMSEs were computed for all such intervals throughout the simulation, and their respective averages, $\overline{RMSE_f}$ and $\overline{RMSE_s}$, served as indices for adaptability to change and estimation accuracy.

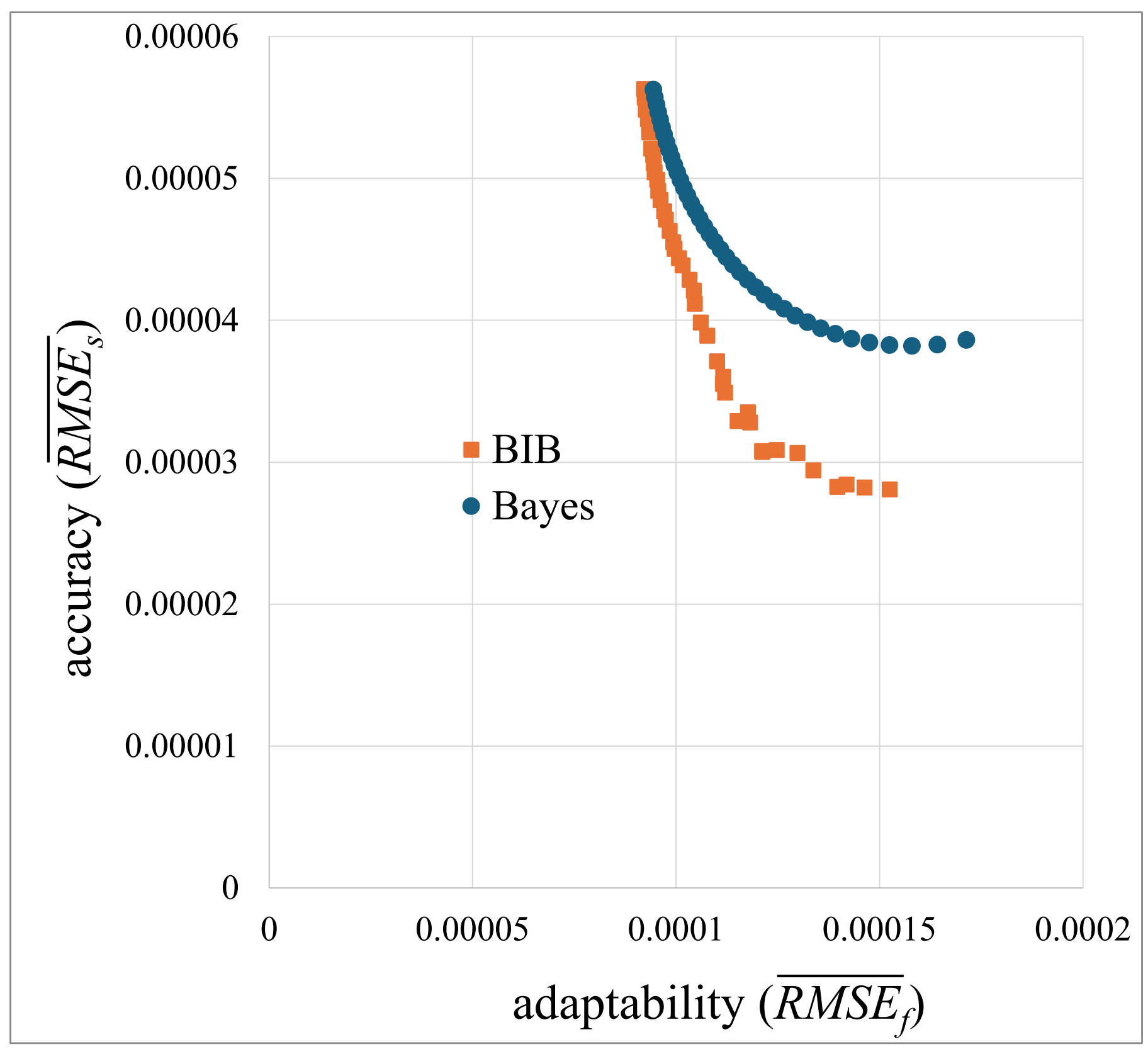

**Fig. 3** Trade-off between adaptability to change and estimation accuracy. The horizontal and vertical axes represent adaptability and accuracy, respectively. Lower values on both axes indicate better performance.

Fig. 3 shows the relationship between adaptability and accuracy for each estimator. Lower values of $\overline{RMSE_f}$ and $\overline{RMSE_s}$ indicate superior performance. The figure shows that β varies from 0.05 to 0.25 in increments of 0.005 for both Bayesian inference and BIB inference. Both methods exhibit a trade-off whereby accuracy corresponds to reduced adaptability. However, the trade-off curve for BIB inference is positioned to the lower left of that for Bayesian inference. The results indicate that BIB inference mitigates the trade-off observed in conventional Bayesian inference.

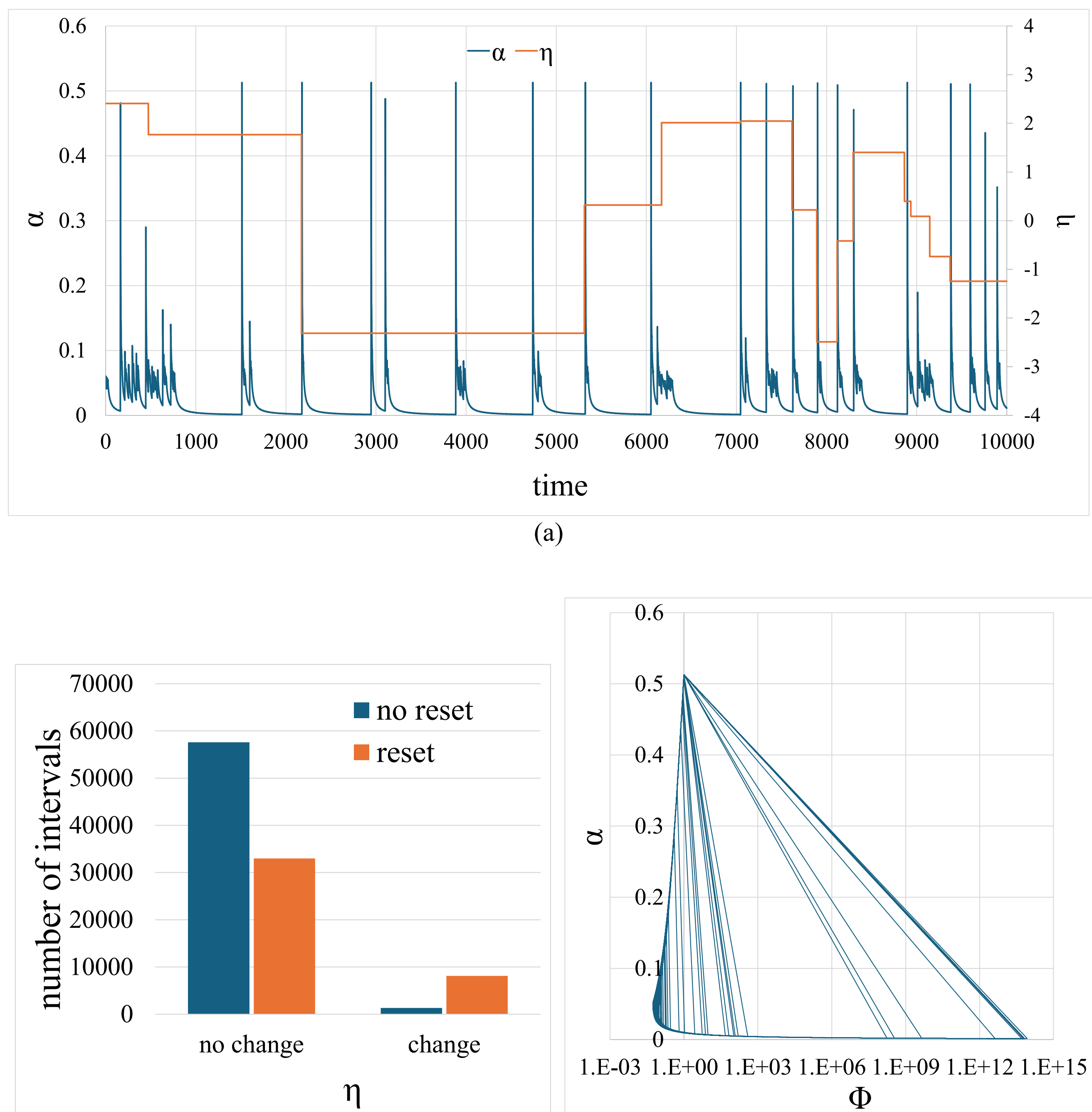

**Fig. 4** Bursts and relaxations in the learning rate during the BIB inference process ( $\beta = 0.05$ ). (a) Time evolution of the learning rate. The mean value η of the data-generating distribution is also shown. (b) Association between changes in η and bursts in the learning rate. (c) Relationship between the learning rate and the variance of the prior distribution. The trajectory connects temporally adjacent data points with lines. The horizontal axis is displayed on a logarithmic scale.

Fig. 4 shows the bursts and relaxations in the learning rate observed during the BIB inference process for $\beta = 0.05$. Fig. 4(a) depicts a representative time series of the learning rate, along with the corresponding time series of the mean $\eta_t$ of the data-generating distribution. Bursts and relaxations in the learning rate occurred intermittently and throughout the time series. Notably, bursts in the learning rate frequently coincided with abrupt changes in $\eta_t$. The learning rate α represents the proportion of external information incorporated into the internal estimate. Hence, it can be interpreted as an indicator of arousal level or sensitivity. Accordingly, the BIB inference mechanism appears to enhance sensitivity in response to significant environmental changes.

The following analyses were performed to examine this relationship. The simulation period (10,000 steps) was divided into 100 intervals. The following conditions were subsequently assessed for each interval:

(i) whether a change in the mean η of the data-generating distribution occurred, and

(ii) whether bursts occurred in the learning rate.

The number of intervals corresponding to each condition was counted. This simulation was executed 1,000 times with different random seeds. Fig. 4(b) shows a summary of the results. A chi-square test of independence revealed a significant association between the change in η and the occurrence of learning rate bursts ($\chi^2\left(1, N = 100000\right) = 8644.79$, $p < 1\times10^{-16}$, $V = 0.29$). Fig. 4(c) depicts the relationship between the learning rate α and the variance Φ of the prior distribution in BIB inference. Although both quantities vary over time, the figure shows their trajectories in phase space. As noted in the previous section, both α and Φ decrease monotonically and eventually converge to their respective asymptotic values, β and $\beta\Sigma_0$, in conventional Bayesian inference. When the discount rate β is large, the learning rate α also increases, resulting in higher sensitivity to incoming data and faster adaptation to abrupt changes. However, the inherent stochasticity of the observed data leads to reduced estimation accuracy. This trade-off between adaptability and accuracy represents a fundamental limitation of Bayesian inference.

In BIB inference, both the learning rate α and the prior variance Φ initially decreased over time, similar to the behavior observed in standard Bayesian inference. However, unlike in the Bayesian case, the learning rate did not converge to β but continued to decline. This continued decrease suppressed fluctuations in the estimate, thereby improving accuracy. Notably, once the learning rate α fell below the discount rate β, the prior variance Φ began to increase and eventually diverged. Abrupt increases were observed during the declining trend of α, corresponding to resets in the inference process; these events appeared as bursts in the learning rate. Consequently, the learning rate did not converge to a fixed value in BIB inference but followed a recurring pattern of bursts and relaxations because of the dynamic interaction between likelihood expansion and resetting.

The time points at which the learning rate α exceeds the threshold β are defined as active states, whereas all other points are classified as rest states.

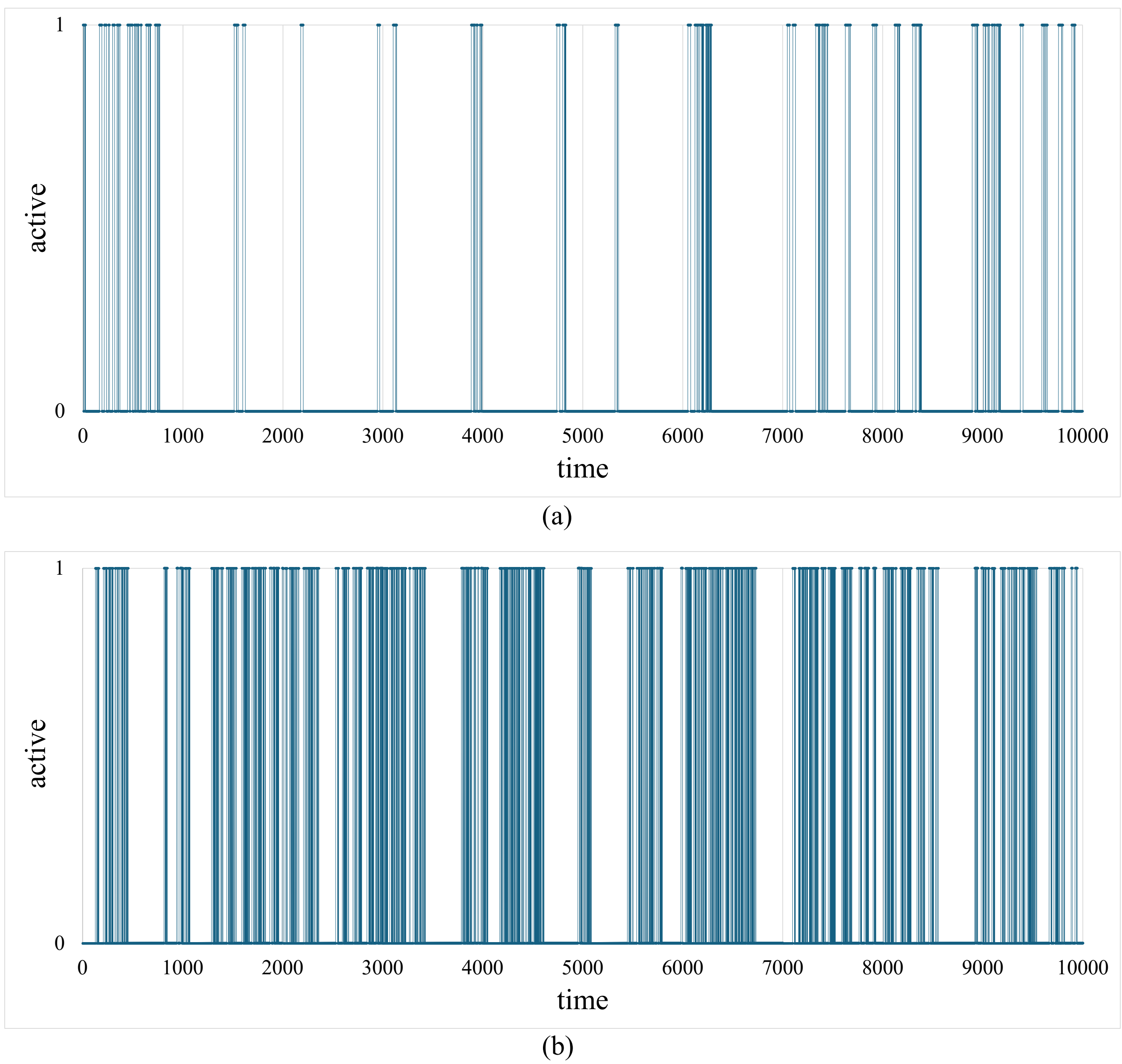

**Fig. 5** Example of the time evolution of active and rest states in BIB inference. Active and rest states are represented as 1 and 0, respectively. (a) $\beta = 0.05$ . (b) $\beta = 0.1$ .

Figs. 5(a) and 5(b) depict time series examples of active and rest states for $\beta = 0.05$ and $\beta = 0.1$ , respectively. In these figures, active states are indicated by 1, and rest states by 0. The rest period $r$ is defined as a continuous interval during which the system remains in the rest state.

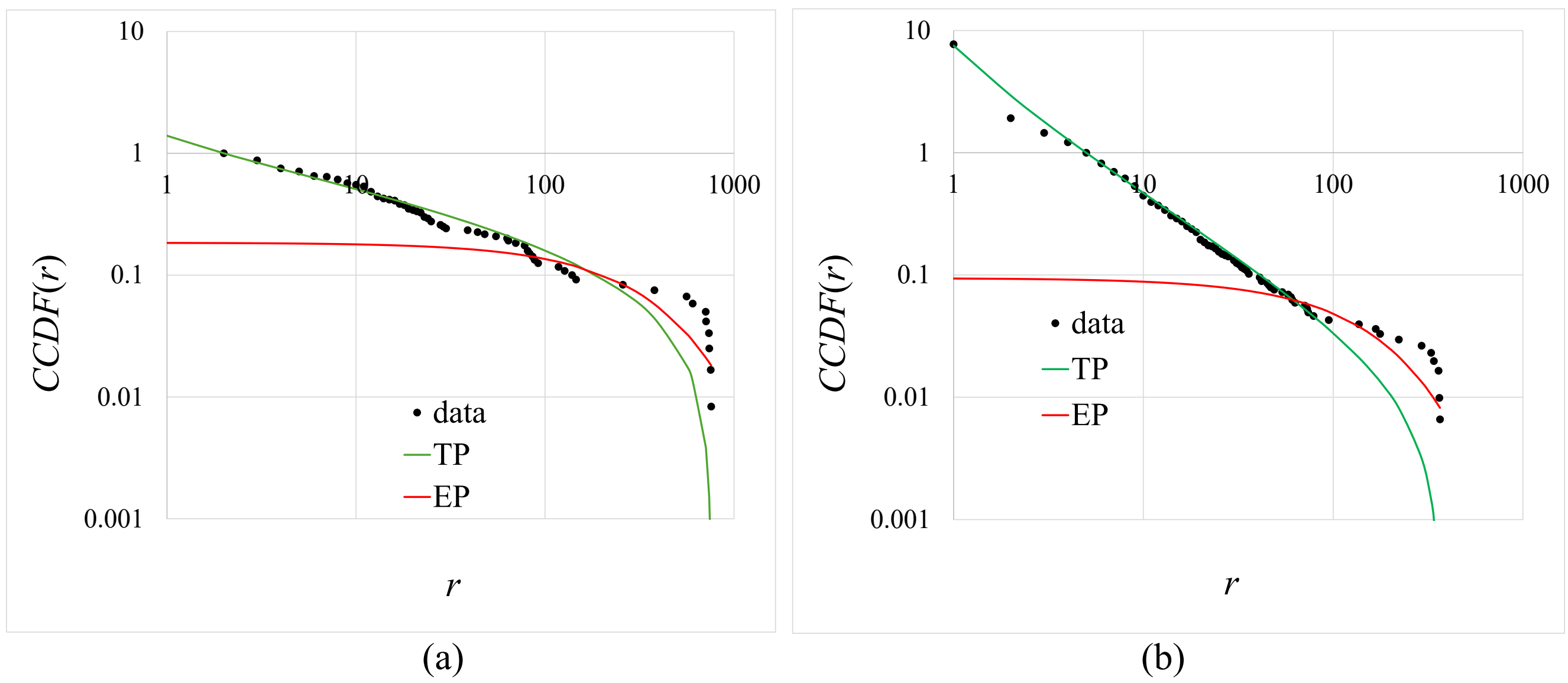

**Fig. 6** CCDF of rest periods corresponding to Figs. 5(a) and 5(b). The fitted results for the TP model (green) and EP model (red) are also shown. (a) CCDF for $\beta = 0.05$. Exponent of TP: 1.29. Fitting range: $[2, 757]$. (b) CCDF for $\beta = 0.1$. Exponent of TP: 2.0. Fitting range: $[5, 366]$.

Figs. 6(a) and 6(b) show the complementary cumulative distribution functions (CCDFs) of the rest period corresponding to Figs. 5(a) and (b), respectively.

When $P(r)$ follows a power-law distribution ($P(r) \propto r^{-\rho}$), the truncated power-law distribution (TP) within the range of $[r_{\min}, r_{\max}]$ is expressed as follows for discrete data:

$$P(r) = \frac{r^{-\rho}}{\zeta(\rho, r_{\min}, r_{\max})}, \quad \zeta(\rho, r_{\min}, r_{\max}) = \sum_{i=r_{\min}}^{r_{\max}} i^{-\rho} \tag{17}$$

$$CCDF(r) = \frac{\zeta(\rho, r, r_{\max})}{\zeta(\rho, r_{\min}, r_{\max})} \tag{18}$$

However, when $P(r)$ follows an exponential distribution ($P(r) \propto e^{-\lambda r}$), the exponential distribution (EP) model is expressed as follows:

$$P(r) = \left(1 - e^{-\lambda}\right) e^{-\lambda(r - r_{\min})} \tag{19}$$

$$CCDF(r) = e^{-\lambda(r - r_{\min})} \tag{20}$$

Substituting $r_{\min}$ for $r$ in Eqs. (18) and (20) yields $CCDF(r_{\min}) = 1$.

Figs. 6(a) and 6(b) depict the results of fitting the TP (green) and EP (red) models to the simulation data. Only the data within the range of $[r_{\min}, r_{\max}]$ from the remaining rest period data were used to optimize the fit. The CCDF value in the CCDF plots was set to one at $r = r_{\min}$. Both axes in Figs. 6(a) and 6(b) are displayed on logarithmic scales, forming double logarithmic plots. This representation applies to all subsequent CCDF graphs. The fitting procedures for the frequency distribution of the rest period *r* to the TP and EP models follow established methods [27–31]. The model parameters were estimated via the maximum likelihood estimation method. The model comparison relied on Akaike information criterion weights, which consider both likelihood and model complexity. Details of the parameter estimation, fitting range $[r_{\min}, r_{\max}]$, and model selection criteria are provided in the Supplementary Information.

As shown in Figs. 6(a) and 6(b), the CCDFs of the rest periods (depicted in Figs. 5(a) and 5(b)) are more accurately fitted by the TP model than by the EP model. At $\beta = 0.05$, out of 1,000 simulations performed with varying random seeds, 179 trials were restricted to a fitting range consisting of only two data points, making reliable estimation of the exponent infeasible. Among the remaining 821 trials, the TP model exhibited a better fit than the EP model in 820 cases. In contrast, when $\beta = 0.1$, 858 out of 1,000 simulations yielded valid fits, and in all of these trials, the TP model provided a superior fit to the EP model.

For the trials that were successfully fitted with the TP model, the estimated scaling exponent ρ was 1.25 ± 0.17 (mean ± SD; n = 820) for $\beta = 0.05$ and 1.70 ± 0.12 (mean ± SD; n = 858) for $\beta = 0.1$.

A t test revealed a statistically significant difference between the two scaling exponents ($t(1676) = -62.94$, $p < 1 \times 10^{-16}$).

### 3.2 Avalanche analysis

We performed avalanche analysis on the data of Simulation 2 to examine the criticality of inference systems. In this study, the subject of the analysis is the absolute value of the estimated variation $|\Delta\theta_t| = |\theta_t - \theta_{t-1}|$. In avalanche analysis, an avalanche is defined as a time interval during which an activity continuously exceeds a certain threshold. Specifically, the threshold is set to the median of $|\Delta\theta_t|$, and an avalanche is identified as a period during which $|\Delta\theta_t|$ consecutively exceeds this threshold. The avalanche duration is denoted by $T$, and the avalanche size $S$ is defined as the total amount of $|\Delta\theta_t|$ during the interval $T$; that is, $S = \sum_{t\in T} |\Delta\theta_t|$. Furthermore, the average size of avalanches with duration $T$ is denoted by $\langle S \rangle_T$.

Avalanche analysis was performed via the following procedure. First, we determined whether the distributions of avalanche size $S$ and duration $T$ follow power-law forms, that is, whether $P(S) \sim S^{-\tau}$ and $P(T) \sim T^{-\tau_t}$ hold. Next, we assessed whether a scaling relationship $\langle S \rangle_T \sim T^{\gamma}$ holds between $\langle S \rangle_T$ and $T$. This fitting was performed using the least squares method. Finally, we verified whether Sethna scaling relation $\gamma = \frac{\tau_t - 1}{\tau - 1}$ holds between these power-law exponents [20, 32]. If all of these are true, it suggests criticality.

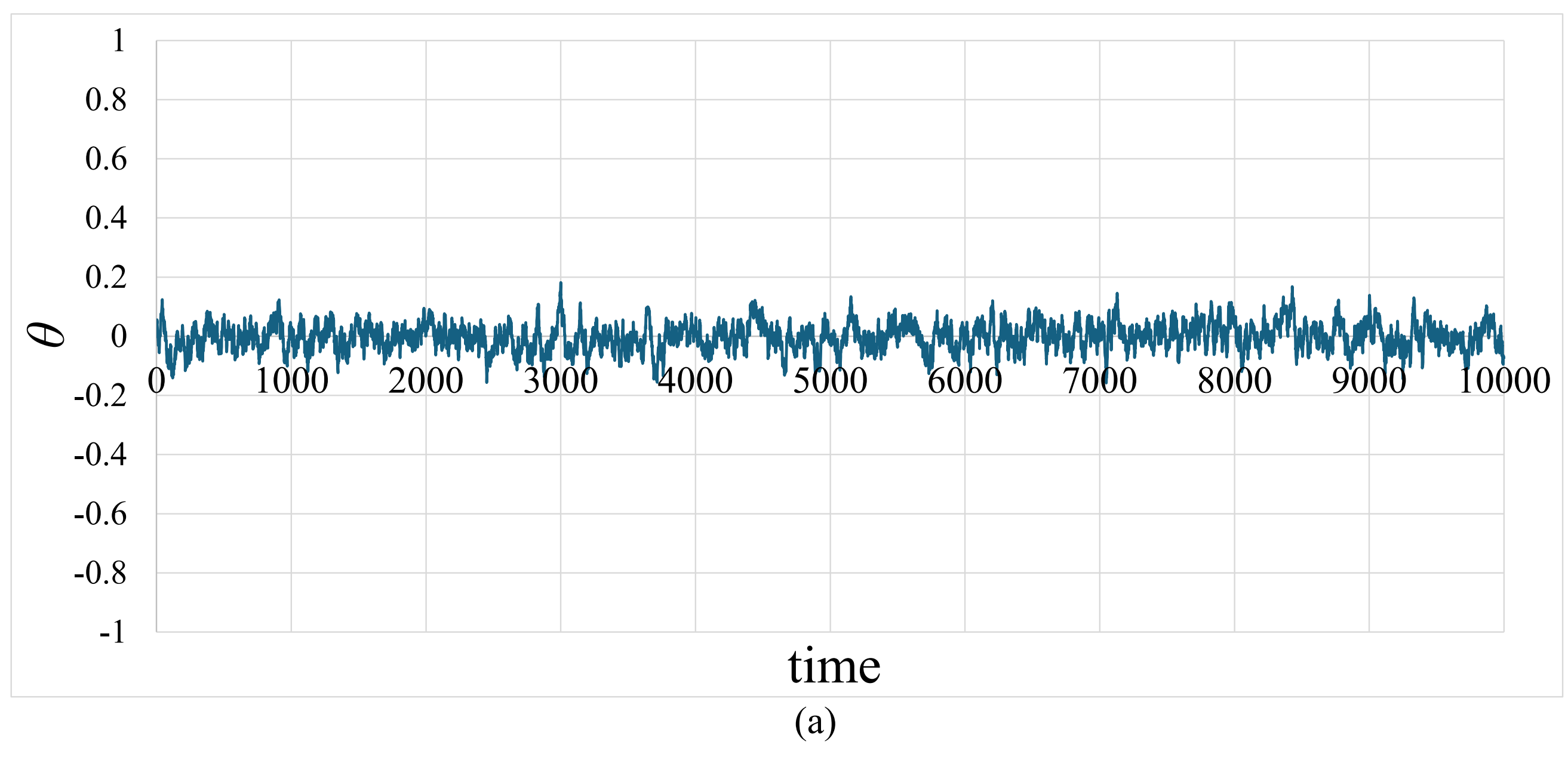

(a)

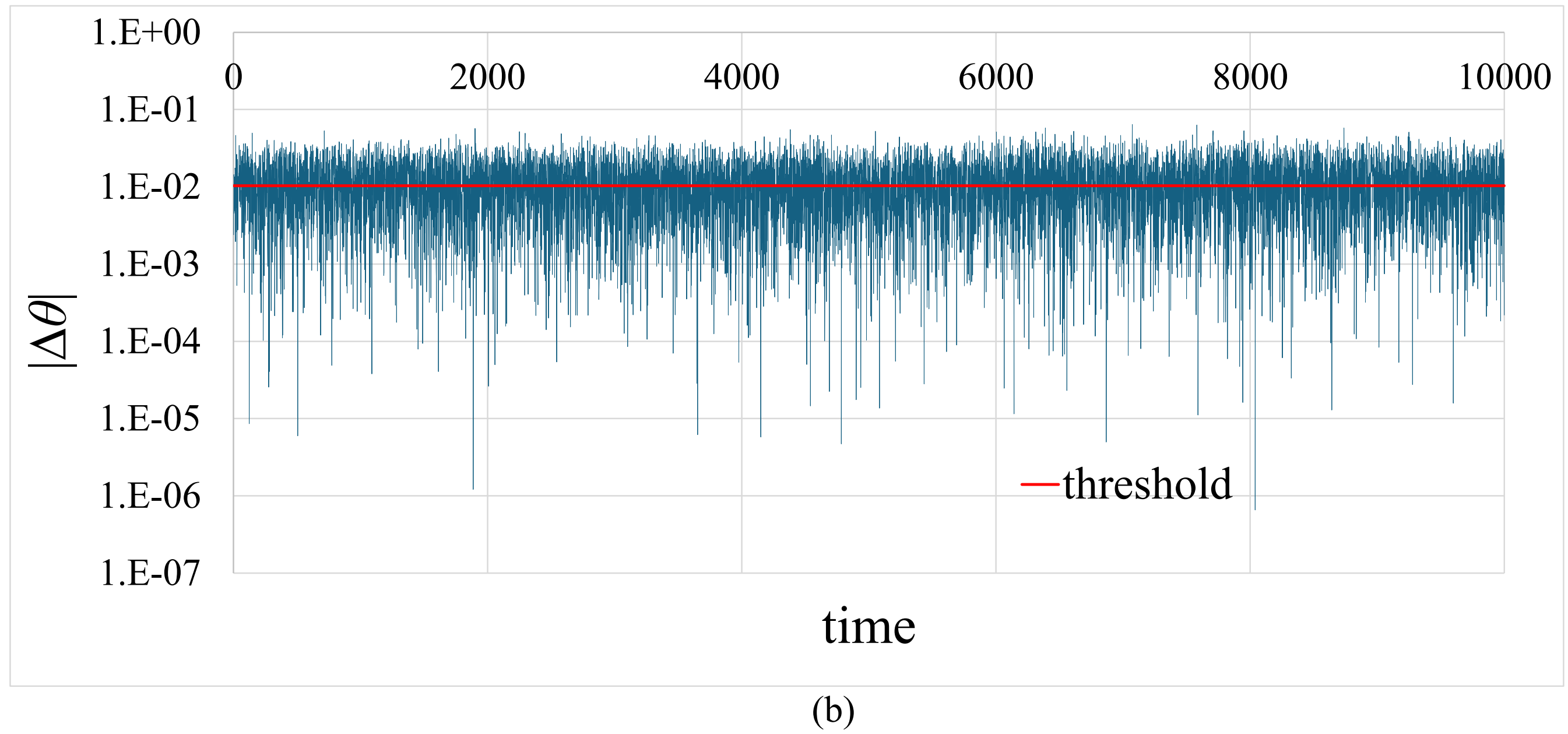

(b)

**Fig. 7** Representative time series of $\theta_t$ and $|\Delta\theta_t|$ using Bayesian inference in simulation 2 ( $\beta = 0.05$ ). (a) Estimate $\theta_t$. The mean value η of the data-generating distribution is zero. (b) Absolute value of the estimated variation $|\Delta\theta_t|$.

Figs. 7(a) and 7(b) depict examples of $\theta_t$ and $|\Delta\theta_t|$ in Bayesian inference, respectively.

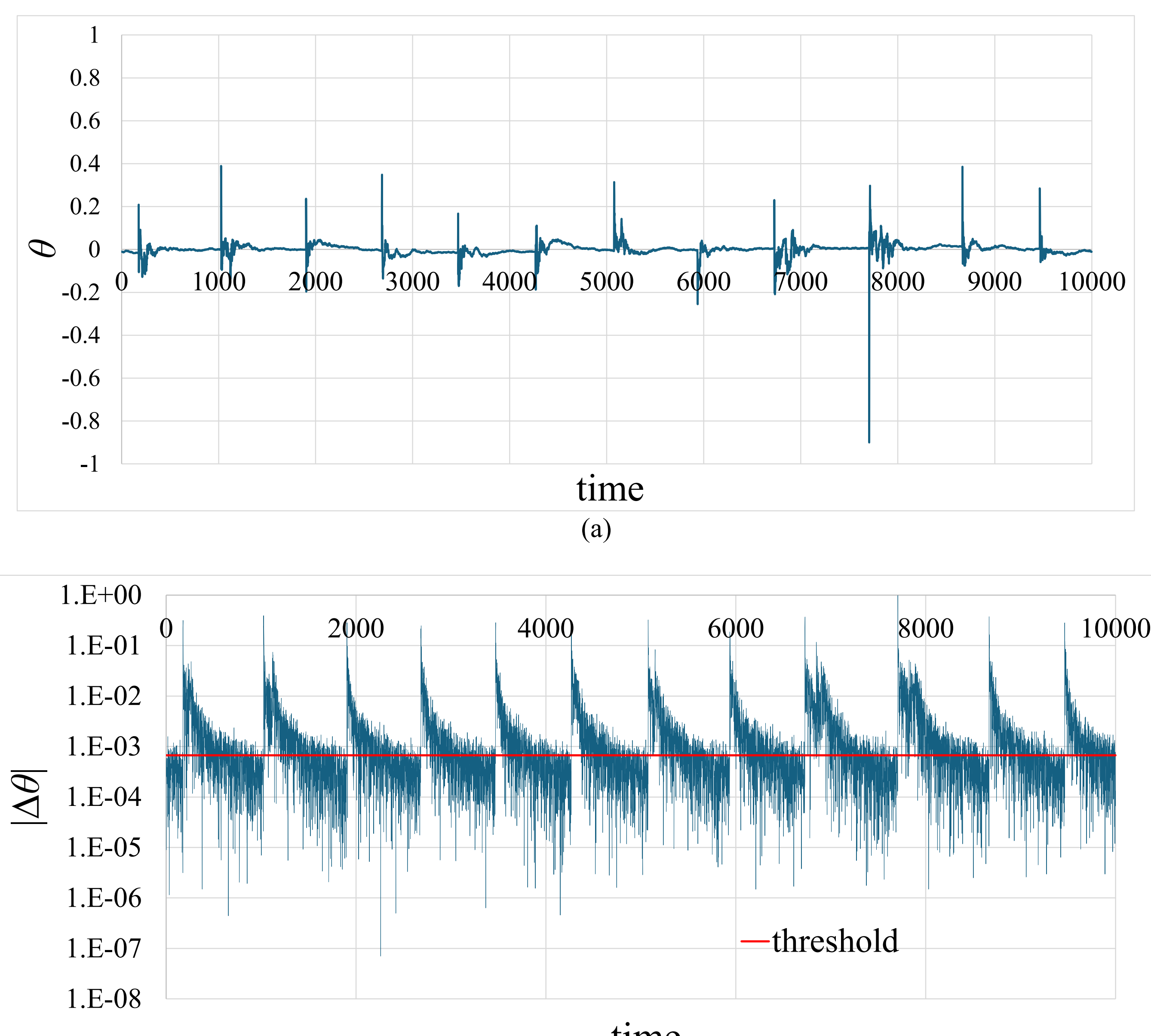

(a)

(b)

**Fig. 8** Representative time series of $\theta_t$ and $|\Delta\theta_t|$ using BIB inference in simulation 2 ( $\beta = 0.05$ ). (a) Estimate $\theta_t$. The mean value η of the data-generating distribution is zero. (b) Absolute value of the estimated variation $|\Delta\theta_t|$.

Figs. 8(a) and 8(b) show examples of estimates $\theta_t$ and $|\Delta\theta_t|$ in BIB inference, respectively. For clarity, the vertical axes in Figs. 7(b) and 8(b) are presented on a logarithmic scale, with the threshold values indicated by red lines in each figure.

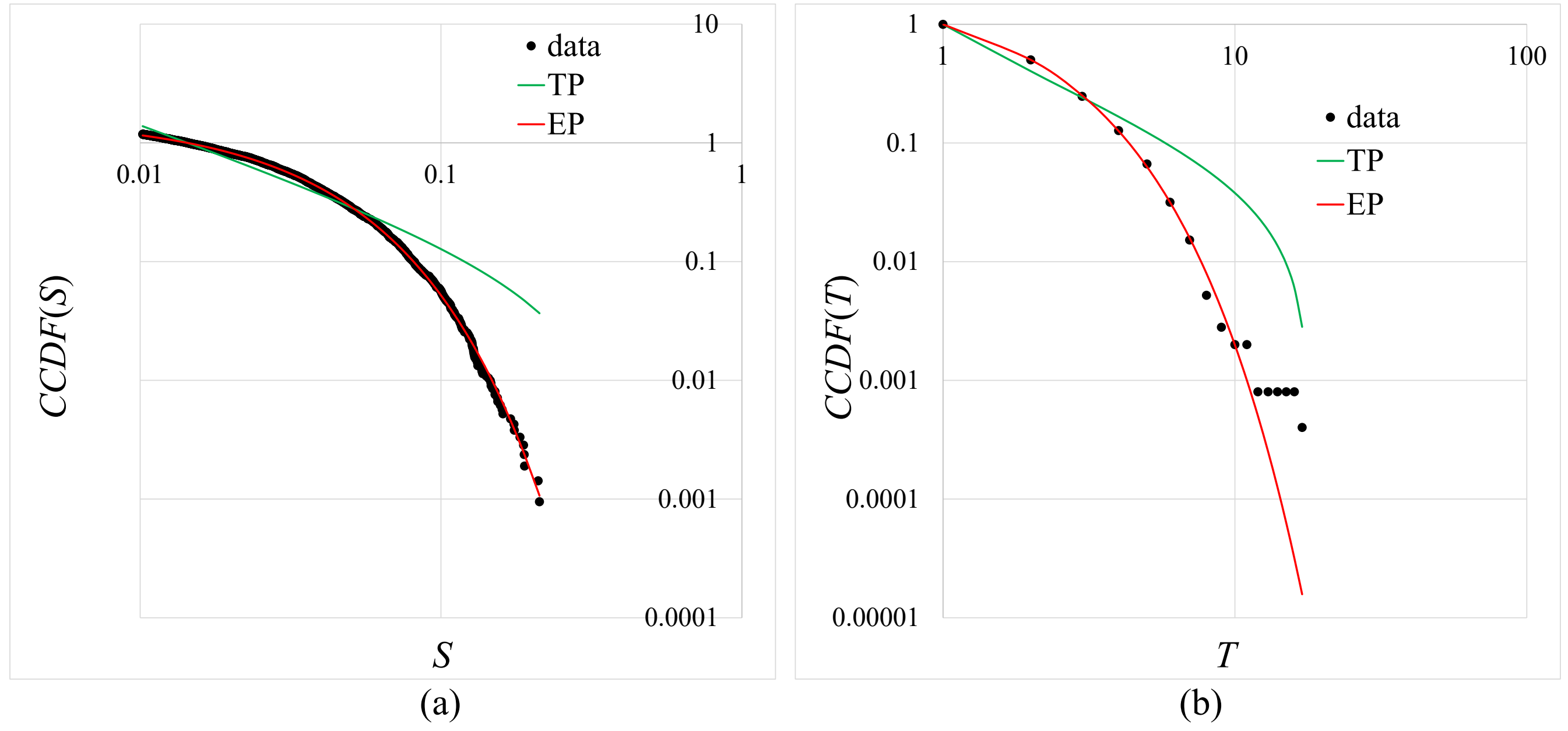

**Fig. 9** Avalanche analyses in Bayesian inference. The fitted results for the TP (green) and EP models (red) are also shown in Figs. 9(a) and 9(b). (a) CCDF of avalanche size. Exponent of EP: 34.54. Fitting range: $[0.0014, 0.37]$. (b) CCDF of avalanche duration. Exponent of EP: 0.69. Fitting range: $[1, 17]$.

Figs. 9(a) and 9(b) show the CCDFs of avalanche size $S$ and avalanche duration $T$ for Bayesian inference, respectively. The plots include the results of fitting the TP (green) and EP (red) models to the data.
Out of 1,000 simulations performed with varying random seeds, 253 trials were restricted to a fitting range consisting of only two data points, making reliable estimation of the exponent infeasible. Among the remaining 747 trials, the EP model consistently provided a better fit to the CCDFs of both $S$ and $T$.

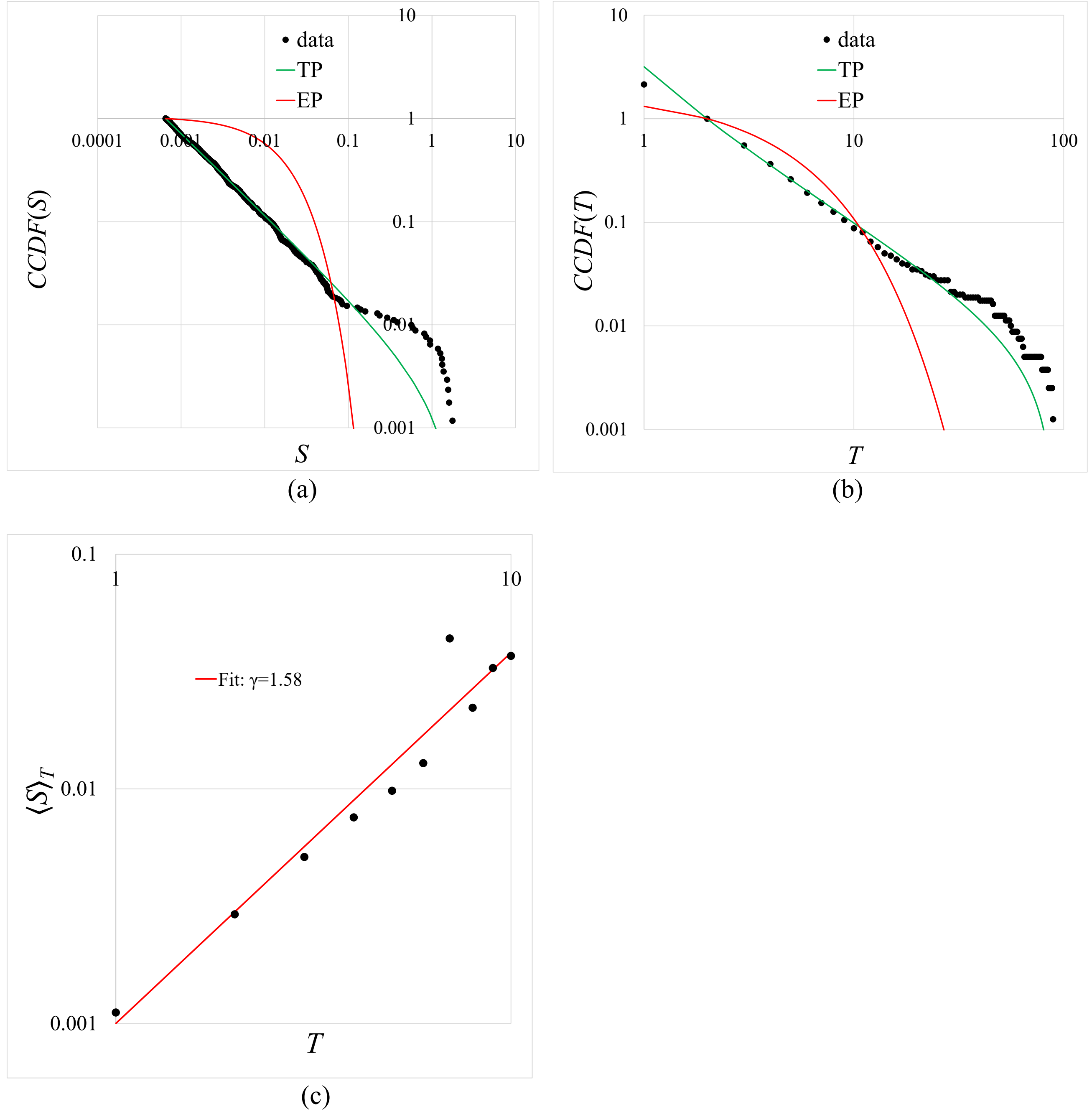

**Fig. 10** Avalanche analyses in BIB inference. The fitted results for the TP (green) and EP models (red) are also shown in Figs. 10(a) and 10(b). (a) CCDF of avalanche size. Exponent of TP: $\tau = 1.79$. Fitting range: $[0.00065, 1.94]$. (b) CCDF of avalanche duration. Exponent of TP: $\tau_t = 2.25$. Fitting range: $[2, 89]$. (c) The scaling relationship between $\langle S \rangle_T$ and $T$ is given by $\langle S \rangle_T \sim T^{\gamma}$. $\gamma = 1.58$. Fitting range: $[1, 10]$.

Figs. 10(a) and 10(b) show the CCDFs of avalanche size $S$ and duration $T$, respectively, for the BIB inference framework. The results of fitting the TP (green) and EP models (red) to the data are also shown. The TP model consistently yielded a better fit to the CCDFs of both $S$ and $T$ across 1,000 simulations for different random seeds. However, in 41 out of 1,000 trials, the fitting range was limited to only two data points, rendering reliable estimation of the power-law exponent infeasible. Consequently, the following analyses were performed using the remaining 959 valid trials.

The scaling exponents $\tau$ and $\tau_t$ were 1.72 ± 0.06 (mean ± SD; n = 959) and 2.14 ± 0.10 (mean ± SD; n = 959), respectively. $\frac{\tau_t - 1}{\tau - 1}$ was 1.58 ± 0.16 (mean ± SD; n = 959). Fig. 10(c) shows the scaling relationship $\langle S \rangle_T \sim T^\gamma$ between $\langle S \rangle_T$ and $T$. Fig. 10(b) shows that the power-law distribution holds approximately within the range of $[1, 10]$; hence, the fitting was performed over this range. The scaling exponent $\gamma$ was 1.60 ± 0.13 (mean ± SD; n = 959).

To confirm the equivalence between $\frac{\tau_t - 1}{\tau - 1}$ and $\gamma$, we employed the Two One-Sided Tests (TOST) procedure with ±5% equivalence bounds based on the mean absolute value of $\gamma$.

Although the paired t test indicated a statistically significant mean difference ($t(958) = -4.12, p = 4.19 \times 10^{-5}$), the TOST showed statistical equivalence ($t_{lower}(958) = 8.46, p = 1.11 \times 10^{-16}, t_{upper}(958) = -16.69, p < 1 \times 10^{-16}$), indicating that the difference was practically negligible. Therefore, $\gamma \approx \frac{\tau_t - 1}{\tau - 1}$ holds true. These results suggest that the BIB inference system is in a critical state [32, 33].

### 3.3 Time domain and frequency domain scaling analysis

To evaluate the presence of LRTC and scale-free behavior in the time series of $|\Delta\theta_t|$, we performed complementary analyses in both the time and frequency domains via DFA and PSA, respectively.

### 3.3.1 DFA

DFA evaluates the root-mean-square fluctuation $F(s)$ of the integrated and detrended signals over various window sizes *s*, resulting in a scaling relationship: $F(s) \sim s^{\omega}$. The resulting power-law exponent $\omega$ is referred to as the DFA exponent and reflects the strength of temporal correlations. An exponent $0.5 < \omega < 1.0$ indicates the presence of LRTC, and values around $\omega \approx 1.0$ are considered indicative of criticality [34].

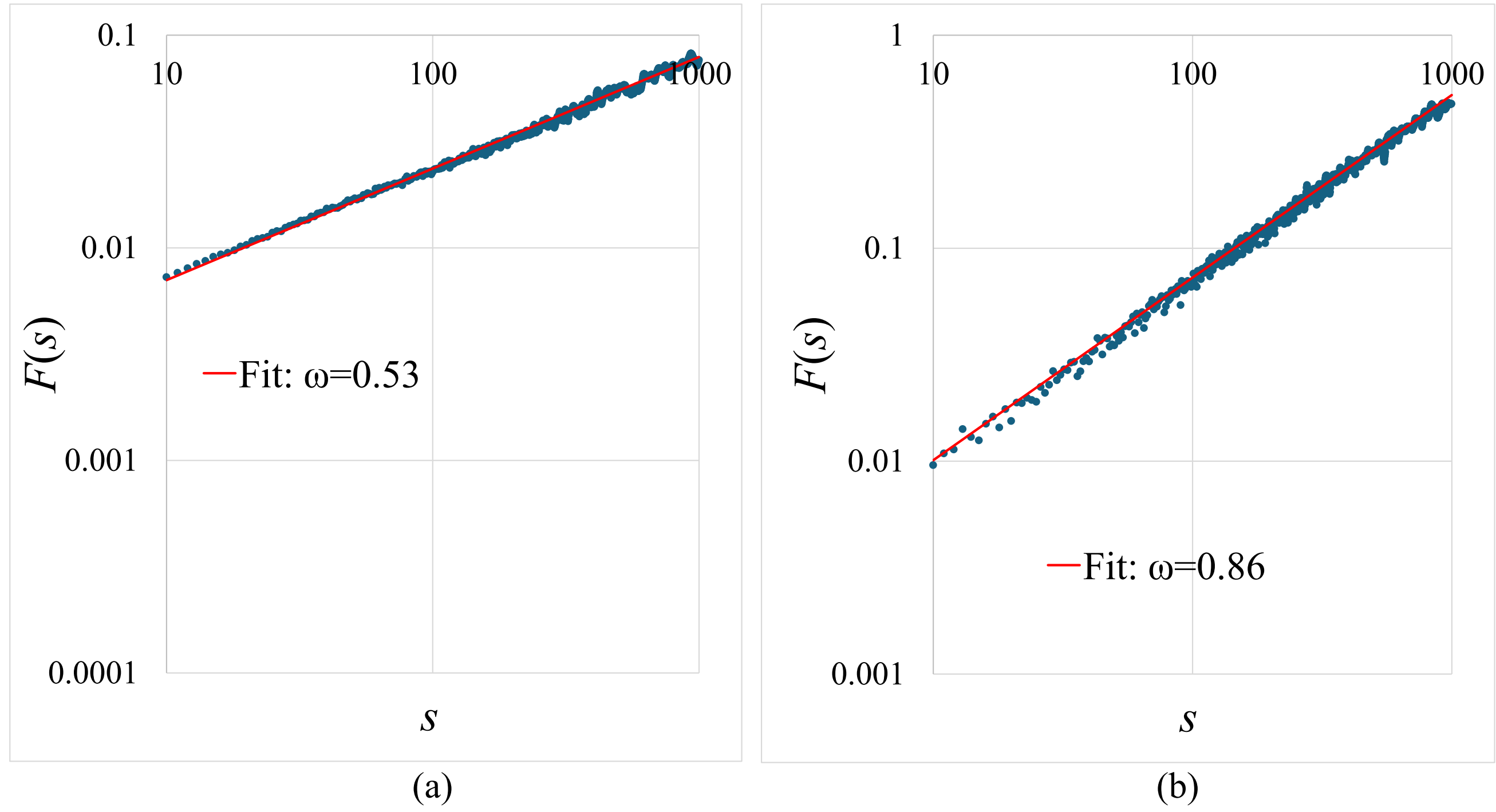

**Fig. 11** Examples of DFA for the time series of $|\Delta\theta_t|$. The scaling relationship between $s$ and $F(s)$ is given by $F(s) \sim s^{\omega}$. The fitting is performed using the least squares method in the range of $[10, 1000]$. (a) DFA exponent $\omega = 0.53$ for Bayesian inference. (b) DFA exponent $\omega = 0.86$ for BIB inference.

Figures 11(a) and 11(b) show representative examples of DFA results for the time series of $|\Delta\theta_t|$ in Bayesian inference and BIB inference, respectively.

The window size $s$ was varied from 10 to 1,000. The DFA exponent $\omega$ was $0.50 \pm 0.03$ (mean ± SD; $n$=747) for Bayesian inference and $0.84 \pm 0.03$ (mean ± SD; $n$=959) for BIB inference. LRTCs were observed only in the BIB inference system. The results suggest that the BIB inference system operates near a critical state.

### 3.3.2 PSA

We applied the fast Fourier transform to the time series data and computed the power spectrum by squaring the amplitudes of the Fourier coefficients. The resulting power spectrum was plotted on a log–log scale, which revealed an approximately linear region in the low-frequency range.

The spectral scaling exponent κ, defined by the power-law relationship between frequency *f* and power $Power(f) \sim f^{-\kappa}$, was estimated by performing linear regression on the log–log plot within the specified frequency range $[0.00024, 0.50]$ (in normalized units).

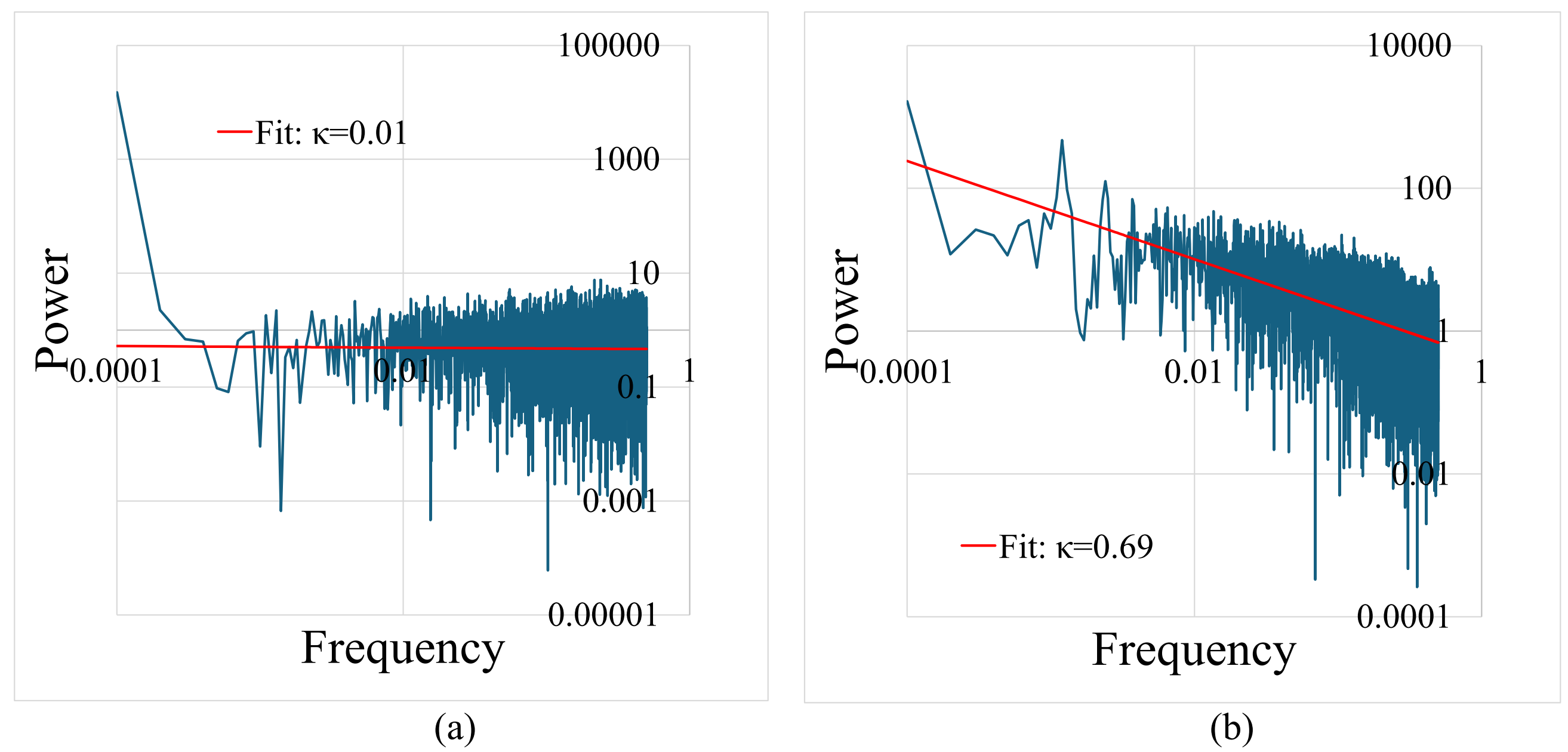

**Fig. 12** Examples of PSA for the time series of $|\Delta\theta_t|$. The scaling relationship between *f* and $Power(f)$ is given by $Power(f) \sim f^{-\kappa}$. The fitting is performed using the least squares method in the range of $[0.00024, 0.50]$. (a) The spectral scaling exponent $\kappa = 0.01$ for Bayesian inference. (b) The spectral scaling exponent $\kappa = 0.69$ for BIB inference.

Figures 12(a) and 12(b) show representative examples of the PSA results for the time series of $|\Delta\theta_t|$ in Bayesian inference and BIB inference, respectively.

The spectral scaling exponent $\kappa$ was 0.035 ± 0.024 (mean ± SD; $n$=747) for Bayesian inference and 0.68 ± 0.12 (mean ± SD; $n$=959) for BIB inference. LRTCs were observed only in the BIB inference system. According to the theoretical results, the spectral exponent $\kappa$ and DFA exponent $\omega$ are related by the equation $\kappa = 2\omega - 1$ [34]. $2\omega - 1$ was 0.68 ± 0.07 (mean ± SD; n = 959) for BIB inference.

To evaluate whether the two variables $2\omega - 1$ and $\kappa$ can be regarded as equivalent, we conducted a TOST procedure with an equivalence margin of ±5% of the mean absolute value of $\kappa$. The result showed statistical equivalence ($t_{lower}(958) = 7.73, p = 1.32 \times 10^{-14}, t_{upper}(958) = -7.61, p = 3.38 \times 10^{-14}$). A paired t test revealed no significant difference between $\kappa$ and $2\omega - 1$ at the 5% significance level ( $t(958) = 0.064,\ p = 0.95$ ).

These results suggest that the theoretical relation $\kappa = 2\omega - 1$ is approximately satisfied in BIB inference. The consistency between the observed values of the DFA and spectral scaling exponents supports the robustness of scale-free dynamics across both the time and frequency domains. Furthermore, the close agreement between these exponents provides strong evidence for the presence of LRTC.

## 4 Discussion and conclusions

This study proposes a novel inference model, BIB inference, by incorporating symmetry bias into Bayesian inference, thereby enabling the concurrent execution of Bayesian and inverse Bayesian updates. However, BIB inference is not a fundamentally distinct form of inference from conventional Bayesian inference. Rather, when the strength of the symmetry bias is zero (i.e., $\beta = 0$), the BIB framework reduces to the standard Bayesian inference.

We applied the proposed model to a task involving the estimation of the mean of a Gaussian distribution generating observed data, where the true mean changed stochastically over time.

First, we confirmed that high learning rates improve adaptability to environmental changes in standard Bayesian inference; however, they reduce estimation accuracy during stable periods owing to excessive sensitivity to incoming data. The outcome reflects an inherent trade-off between adaptability and accuracy. However, the BIB inference mitigates this trade-off, achieving both rapid responsiveness to abrupt changes and high accuracy during stable intervals.

This improvement results from the distinct dynamic properties of the learning rate in Bayesian and BIB inferences. In standard Bayesian inference, the accumulation of observational data increases confidence in the estimate, causing the learning rate to gradually decline and eventually converge to a constant value. However, inverse Bayesian updates produce intermittent bursts and relaxations in the learning rate over time.

As shown in Fig. 4(c), BIB inference exhibits a seemingly paradoxical regime: after a certain point, the posterior (next-step prior) variance Φ expands over time, whereas the learning rate α decays toward zero until a reset occurs. Although greater uncertainty might be expected to increase learning, the learning rate $\alpha_{t+1} = \frac{\Phi_t}{\Phi_t + (1-\beta)\Sigma_t}$ reflects the balance between state-side uncertainty (Φ) and observational uncertainty (Σ). Thus, even as Φ increases, α can still decline when Σ increases faster. In our implementation, inverse Bayesian updates endogenously inflate Σ, and intermittent resets produce a characteristic cycle: α burst relaxation.

Previous studies have consistently demonstrated that adaptive learning rates exhibit a burst–relaxation dynamic in response to environmental uncertainty. For example, Behrens et al. [35] reported that activity in the anterior cingulate cortex encodes environmental volatility, with both neural activity and learning rates transiently increasing after unexpected changes before returning to baseline. Computational modeling further supports this view: Nassar et al. [36] introduced an approximately Bayesian delta-rule model in which learning rates rise sharply at change points, governed by change-point probability and relative uncertainty, and then decrease as stability is regained. Mathys et al. [37] extended this framework to the hierarchical Gaussian filter, a hierarchical Bayesian model that explicitly represents volatility at multiple levels of a generative process and adapts learning accordingly. Complementing these computational accounts, Jepma et al. [38] provided pharmacological and EEG evidence that catecholaminergic neuromodulation—particularly noradrenaline—causally regulates trial-by-trial adjustments of the learning rate in dynamic environments.

The latent cause framework suggests that, in response to environmental change, learners internally generate new contextual representations, thereby producing fluctuations analogous to the resetting mechanism in the BIB framework and inducing bursts in the learning rate [39, 40].

In addition to external volatility, internal mechanisms have also been proposed to generate similar dynamics. The adaptive gain theory of the locus coeruleus–noradrenaline (LC–NE) system posits that spontaneous fluctuations in arousal modulate the precision weighting of prediction errors, producing transient increases and subsequent relaxations in the learning rate even in the absence of environmental change [41].

Within this line of research, our BIB inference model provides a natural extension. By integrating inverse Bayesian updates with conventional Bayesian inference, the BIB framework enables dynamic restructuring of the hypothesis space, producing spontaneous burst–relaxation cycles of the learning rate not only at external change points but also through intrinsic resetting. In contrast to earlier models that attributed burst–relaxation exclusively to external volatility, the BIB model provides a more general and mechanistic account in which adaptive bursts emerge intrinsically from the dynamics of inference itself.

Interpreting the learning rate as an index of sensitivity to external stimuli reveals a compelling parallel with biological survival strategies. In stable environments, conserving energy through rest is advantageous, whereas alertness is necessary for responding to unexpected changes. The learning rate in standard

Bayesian inference is fixed; hence, high sensitivity incurs a cost to rest, and prioritizing rest reduces environmental responsiveness, rendering the trade-off unavoidable. Certain animals, such as birds and aquatic mammals, have evolved unihemispheric slow-wave sleep (USWS), which enables one hemisphere of the brain to sleep while the other remains awake, thereby balancing energy conservation with vigilance [42]. In BIB inference, a dynamic alteration between the active and resting phases emerges naturally during belief updating. When unexpected changes occur, the learning rate spontaneously increases, triggering a transition to a high-sensitivity state. These findings suggest that BIB inference involves a flexible mechanism capable of balancing two critical requirements for biological agents: energy conservation and rapid adaptation to environmental variability.

We analyzed the distributions of the remaining periods in BIB inference. They consistently exhibited power-law distributions. Although the scaling exponent changed with the strength of the symmetry bias, the scaling structure remained robust, indicating the universality of this dynamic pattern in the proposed model. Time intervals between behavioral events follow power-law distributions from a biological perspective. For example, the intervals between actions such as sending emails, accessing websites, and posting on social media typically exhibit heavy-tailed power-law characteristics rather than simple exponential patterns [43]. This phenomenon has been attributed to priority-based queuing models, which assume that task execution is governed by priority selection rather than random choice, producing scale-free interevent intervals. However, the universality of this explanation remains under debate [44]. Pfister and Ghosh [45] extended the priority-based queuing model to consider the statistical properties of time intervals and bursty behavior observed in human smartphone usage, such as screen-touch patterns.

Additionally, notable studies have reported associations between mental disorders and scale-free behavioral dynamics. For example, in experiments comparing individuals with major depressive disorder and healthy controls, both groups displayed power-law distributions in resting period durations. However, the scaling exponent differed on the basis of the presence of disorder [46]. Studies on *Drosophila melanogaster* have shown that dopaminergic activity modulates the duration of rest periods, yet the distribution consistently retains a power-law structure [47]. Similarly, whereas insulin signaling and nutritional conditions influence the temporal characteristics of activity, both the active and resting periods follow frequency distributions resembling power laws in *Caenorhabditis elegans* [48]. These findings suggest that even organisms with simple nervous systems can exhibit spontaneous scale-free behavioral patterns that are modulated by internal metabolic states and external environmental conditions.

The emergence of scale-free structures is frequently interpreted through the framework of self-organized criticality (SOC). SOC describes a phenomenon wherein a system spontaneously evolves toward a critical state—at the boundary between order and disorder—without requiring detailed fine-tuned external control, thereby exhibiting scale-free behavior [17].

In this study, we applied avalanche analysis, DFA, and PSA to determine whether standard Bayesian inference and the BIB framework operate near a critical state. The results show that criticality occurs exclusively in the BIB system, suggesting that BIB inference uniquely facilitates the coexistence of computational efficiency and critical dynamics.

This interpretation aligns with previous theoretical and empirical studies indicating that neural criticality confers functional advantages, including optimal information transmission, heightened sensitivity to external inputs, and enhanced computational capacity [18, 19].

A central parameter in the BIB model is β, which regulates the strength of the symmetry bias. In conventional Bayesian updates, β functions as a discounting factor that modulates the influence of prior beliefs. In contrast, during inverse Bayesian updates, β governs the expansion of the likelihood variance. Therefore, β plays a dual role in shaping inference dynamics within the BIB framework.

The fundamental distinction between BIB inference and standard Bayesian inference lies in the incorporation of inverse Bayesian updates. Standard Bayesian inference with a discounting mechanism can be regarded as a special case of the BIB framework in which the variance expansion function of β is inactive.

The inverse Bayesian updating process emerges naturally through the integration of symmetry bias into conventional Bayesian inference. However, the specific real-world physical phenomenon to which this process corresponds remains unclear. Moreover, experimental evidence is needed to establish whether inverse Bayesian updating, in addition to standard Bayesian updating, is genuinely involved in the reasoning processes of humans and other biological systems. Clarifying this question represents a critical direction for future research.

**Declaration of generative AI and AI-assisted technologies in the manuscript preparation process**

During the preparation of this work the author(s) used ChatGPT in order to check the English. After using this tool/service, the author(s) reviewed and edited the content as needed and take(s) full responsibility for the content of the published article.

## References

[1] Friston KJ, Kilner J, Harrison L. A free energy principle for the brain. J Physiol Paris. 2006;100:70–87. https://doi.org/10.1016/j.jphysparis.2006.10.001.

[2] Friston KJ, Kiebel S. Cortical circuits for perceptual inference. Neural Netw. 2009;22:1093–104. https://doi.org/10.1016/j.neunet.2009.07.023.

[3] Hohwy J. The predictive mind. Oxford University Press; 2013. https://doi.org/10.1093/acprof:oso/9780199682737.001.0001.

[4] Friston KJ. The free-energy principle: a unified brain theory? Nat Rev Neurosci. 2010;11:127–38. https://doi.org/10.1038/nrn2787.

[5] Dehaene S. Consciousness and the brain: deciphering how the brain codes our thoughts. New York: Viking Press; 2014.

[6] Chater N, Oaksford M, editors. The probabilistic mind: prospects for Bayesian cognitive science. England: Oxford University Press; 2008.

[7] Sidman M, Tailby W. Conditional discrimination vs. matching-to-sample: an expansion of the testing paradigm. J Exp Anal Behav. 1982;37:5–22. https://doi.org/10.1901/jeab.1982.37-5.

[8] Shinohara S, Taguchi R, Katsurada K, Nitta T. A model of belief formation based on causality and application to N-armed bandit problem. Trans Jpn Soc Artif Intell. 2007;22:58–68. https://doi.org/10.1527/tjsai.22.58.

[9] Shinohara S, Manome N, Suzuki K, Chung UI, Takahashi T, Gunji PY et al. Extended Bayesian inference incorporating symmetry bias. Biosystems. 2020;190:104104. https://doi.org/10.1016/j.biosystems.2020.104104.

[10] Markman EM, Wachtel GF. Children's use of mutual exclusivity to constrain the meanings of words. Cogn Psychol. 1988;20:121–57. https://doi.org/10.1016/0010-0285(88)90017-5.

[11] Hattori M, Oaksford M. Adaptive non-interventional heuristics for covariation detection in causal induction: model comparison and rational analysis. Cogn Sci. 2007;31:765–814. https://doi.org/10.1080/03640210701530755.

[12] Shinohara S, Manome N, Suzuki K, Chung UI, Takahashi T, Okamoto H et al. A new method of Bayesian causal inference in non-stationary environments. PLOS One. 2020;15:e0233559. https://doi.org/10.1371/journal.pone.0233559.

[13] Gunji Y-P, Shinohara S, Haruna T, Basios V. Inverse Bayesian inference as a key of consciousness featuring a macroscopic quantum logical structure. Biosystems. 2017;152:44–65. https://doi.org/10.1016/j.biosystems.2016.12.003.

[14] Gunji Y-P, Murakami H, Tomaru T, Basios V. Inverse Bayesian inference in swarming behaviour of soldier crabs. Philos Trans A Math Phys Eng Sci. 2018;376:20170370. https://doi.org/10.1098/rsta.2017.0370.

[15] Horry Y, Yoshinari A, Nakamoto Y, Gunji Y-P. Modeling of decision-making process for moving straight using inverse Bayesian inference. Biosystems. 2018;163:70–81. https://doi.org/10.1016/j.biosystems.2017.12.006.

[16] Gunji YP, Kawai T, Murakami H, Tomaru T, Minoura M, Shinohara S. Lévy walk in swarm models based on Bayesian and inverse Bayesian inference. Comput Struct Biotechnol J. 2021;19:247–60. https://doi.org/10.1016/j.csbj.2020.11.045.

[17] Bak P, Tang C, Wiesenfeld K. Self-organized criticality: an explanation of the 1/f noise. Phys Rev Lett. 1987;59:381–4. https://doi.org/10.1103/PhysRevLett.59.381.

[18] Beggs JM, Plenz D. Neuronal avalanches in neocortical circuits. J Neurosci. 2003;23:11167–77. https://doi.org/10.1523/JNEUROSCI.23-35-11167.2003.

[19] Shew WL, Plenz D. The functional benefits of criticality in the cortex. Neuroscientist. 2013;19:88–100. https://doi.org/10.1177/1073858412445487.

[20] Sethna JP, Dahmen KA, Myers CR. Crackling noise. Nature. 2001 Mar 8;410(6825):242-50. doi: 10.1038/35065675. PMID: 11258379.

[21] Friedman N, Ito S, Brinkman BAW, Shimono M, DeVille REL, Dahmen KA et al. Universal critical dynamics in high-resolution neuronal avalanche data. Phys Rev Lett. 2012;108:208102. https://doi.org/10.1103/PhysRevLett.108.208102.

[22] Shriki O, Alstott J, Carver F, Holroyd T, Henson RNA, Smith ML et al. Neuronal avalanches in the resting MEG of the human brain. J Neurosci. 2013;33:7079–90. https://doi.org/10.1523/JNEUROSCI.4286-12.2013.

[23] Yu S, Ribeiro TL, Meisel C, Chou S, Mitz A, Saunders R et al. Maintained avalanche dynamics during task-induced changes of neuronal activity in nonhuman primates. Elife. 2017;6:e27119. https://doi.org/10.7554/eLife.27119.

[24] Palva JM, Zhigalov A, Hirvonen J, Korhonen O, Linkenkaer-Hansen K, Palva S. Neuronal long-range temporal correlations and avalanche dynamics are correlated with behavioral scaling laws. Proc Natl Acad Sci U S A. 2013;110:3585–90. https://doi.org/10.1073/pnas.1216855110.

[25] http://www.mingw.org/ (2025). Accessed 2 May 2025

[26] https://www.qt.io/ (2025). Accessed 2 May 2025

[27] Jansen VAA, Mashanova A, Petrovskii S. Comment on "Lévy walks evolve through interaction between movement and environmental complexity". Science. 2012;335:918; author reply 918. https://doi.org/10.1126/science.1215747.

[28] Humphries NE, Weimerskirch H, Queiroz N, Southall EJ, Sims DW. Foraging success of biological Lévy flights recorded in situ. Proc Natl Acad Sci U S A. 2012;109:7169–74. https://doi.org/10.1073/pnas.1121201109.

[29] White EP, Enquist BJ, Green JL: on estimating the exponent of power-law frequency distributions Ecology. Erratum. In: Ecology 89. 2008. 10.1890/07-1288.1;89:905–12. https://doi.org/10.1890/07-1288.1.

[30] Clauset A, Shalizi CR, Newman MEJ. Newman Power-law distributions in empirical data in Siam. SIAM Rev. 2009;51:661–703. https://doi.org/10.1137/070710111.

[31] Edwards AM, Phillips RA, Watkins NW, Freeman MP, Murphy EJ, Afanasyev V et al. Revisiting Lévy flight search patterns of wandering albatrosses, bumblebees and deer. Nature. 2007;449:1044–8. https://doi.org/10.1038/nature06199.

[32] Carvalho TTA, Fontenele AJ, Girardi-Schappo M, Feliciano T, Aguiar LAA, Silva TPL et al. Subsampled directed-percolation models explain scaling relations experimentally observed in the brain. Front Neural Circuits. 2020;14:576727. https://doi.org/10.3389/fncir.2020.576727.

[33] Zimmern V. Why brain criticality is clinically relevant: A scoping review. Front Neural Circuits. 2020;14:54. https://doi.org/10.3389/fncir.2020.00054.

[34] Heneghan C, McDarby G. Establishing the relation between detrended fluctuation analysis and power spectral density analysis for stochastic processes. Phys Rev E. 2000;62:6103-10. https://doi.org/10.1103/physreve.62.6103.

[35] Behrens TEJ, Woolrich MW, Walton ME, Rushworth MFS, Learning the value of information in an uncertain world, Nat. Neurosci. 10 (2007) 1214–1221. https://doi.org/10.1038/nn1954.

[36] Nassar MR, Wilson RC, Heasly B, Gold JI. An approximately Bayesian delta-rule model explains the dynamics of belief updating in a changing environment. J Neurosci. 2010 Sep 15;30(37):12366-78. https://doi.org/10.1523/JNEUROSCI.0822-10.2010.

[37] Mathys C, Daunizeau J, Friston KJ, Stephan KE. A bayesian foundation for individual learning under uncertainty. Front Hum Neurosci. 2011 May 2;5:39. https://doi.org/10.3389/fnhum.2011.00039.

[38] Jepma M, Murphy PR, Nassar MR, Rangel-Gomez M, Meeter M, Nieuwenhuis S. Catecholaminergic Regulation of Learning Rate in a Dynamic Environment. PLoS Comput Biol. 2016 Oct 28;12(10):e1005171. https://doi.org/10.1371/journal.pcbi.1005171.

[39] Gershman SJ, Blei DM, Niv Y. Context, learning, and extinction. Psychol Rev. 2010;117(1):197-209. https://doi.org/10.1037/a0017808.

[40] Gershman SJ, Niv Y. Exploring a latent cause theory of classical conditioning. Learn. Behav. 2012; 40 (3) : 255–268. https://doi.org/10.3758/s13420-012-0080-8.

[41] Aston-Jones G, Cohen JD. An integrative theory of locus coeruleus-norepinephrine function: adaptive gain and optimal performance. Annu Rev Neurosci. 2005;28:403-50. https://doi.org/10.1146/annurev.neuro.28.061604.135709.

[42] Rattenborg NC, Voirin B, Cruz SM, Tisdale R, Dell'Omo G, Lipp HP et al. Evidence that birds sleep in mid-flight. Nat Commun. 2016;7:12468. https://doi.org/10.1038/ncomms12468.

[43] Barabási AL. The origin of bursts and heavy tails in human dynamics. Nature. 2005;435:207–11. https://doi.org/10.1038/nature03459.

[44] Malmgren RD, Stouffer DB, Motter AE, Amaral LAN. A Poissonian explanation for heavy tails in e-mail communication. Proc Natl Acad Sci U S A. 2008;105:18153–8. https://doi.org/10.1073/pnas.0800332105.

[45] Pfister JP, Ghosh A. Generalized priority-based model for smartphone screen touches. Phys Rev E. 2020;102(1-1):012307. https://doi.org/10.1103/PhysRevE.102.012307.

[46] Nakamura T, Kiyono K, Yoshiuchi K, Nakahara R, Struzik ZR, Yamamoto Y. Universal scaling law in human behavioral organization. Phys Rev Lett. 2007;99:138103. https://doi.org/10.1103/PhysRevLett.99.138103.

[47] Ueno T, Masuda N, Kume S, Kume K. Dopamine modulates the rest period length without perturbation of its power law distribution in Drosophila melanogaster. PLOS One. 2012;7:e32007. https://doi.org/10.1371/journal.pone.0032007.

[48] Arata Y, Shiga I, Ikeda Y, Jurica P, Kimura H, Kiyono K et al. Insulin signaling shapes fractal scaling of C. elegans behavior. Sci Rep. 2022;12:10481. https://doi.org/10.1038/s41598-022-13022-6.

## Funding

This work was supported by JSPS KAKENHI [grant numbers JP21K12009 and 25K15214].

## Data availability statement

The code used to perform the simulations in this study is available on GitHub at: https://github.com/shinoharaken/BIBinference.

## Competing Interests

The authors declare no competing interests.

# Supplementary Information for

# Adaptive Inference through Bayesian and Inverse Bayesian Inference with Symmetry Bias in Nonstationary Environments

## Fitting to simulation data

Here, a method is described to fit the frequency distribution of the data observed via simulation to the truncated power-law distribution (TP) model and the exponential distribution (EP) model. Here, the case in which the data take continuous values is discussed. For discrete values, please refer to [1].

## Fitting to TP

Here, a method is described to fit the frequency distribution of the step length $l$ observed by simulation to the TP. The method is based on previous studies [2–6]. Specifically, the aim is to determine the minimum $\hat{l}_{\min}$ and maximum values $\hat{l}_{\max}$ of the observed data to be fitted to the TP model and the exponent $\hat{\eta}$ of the TP model that best fits the data in the range $\hat{l}_{\min} \le l \le \hat{l}_{\max}$. First, $\hat{l}_{\max}$ is the longest step length of the observation data. Next, the calculation method $\hat{l}_{\min}$ is described. In the case of a continuous distribution, TP in the range $l_{\min} \le l \le l_{\max}$ is expressed by the following formula:

$$p\left(l;\eta,l_{\min},l_{\max}\right)=\frac{\eta-1}{l_{\min}^{1-\eta}-l_{\max}^{1-\eta}}l^{-\eta} \tag{S21}$$

The complementary cumulative distribution function (CCDF) of $p\left(l;\eta,l_{\min},l_{\max}\right)$ is expressed in the following equation.

$$P(l;\eta,l_{\min},l_{\max}) = 1-\frac{l_{\min}^{1-\eta}-l^{1-\eta}}{l_{\min}^{1-\eta}-l_{\max}^{1-\eta}} = \frac{l^{1-\eta}-l_{\max}^{1-\eta}}{l_{\min}^{1-\eta}-l_{\max}^{1-\eta}} \quad \text{(S22)}$$

If the observed data in the range of $l_{\min} \le l \le l_{\max}$ are $\{l_1, l_2, \cdots, l_n\}$, then the log-likelihood of these data for TP is calculated using Equation (S1) as follows:

$$L(\eta;l_{\min},l_{\max}) = \sum_{i=1}^{n} \ln p(l_i;\eta,l_{\min},l_{\max}) = n\left(\ln(\eta-1)-\ln\left(l_{\min}^{1-\eta}-l_{\max}^{1-\eta}\right)\right)-\eta\sum_{i=1}^{n}\ln l_i \quad \text{(S23)}$$

The exponent $\hat{\eta}(l_{\min},l_{\max})$ of the TP model that best fits the data in the range $l_{\min} \le l \le l_{\max}$ is $\eta$, which maximizes $L(\eta;l_{\min},l_{\max})$. Specifically, $\eta$ is varied from 0.5 to 3.5, in increments of 0.01, to obtain $\hat{\eta}(l_{\min},l_{\max})$, which numerically maximizes Equation (S3).

The authors introduced the Kolmogorov–Smirnov statistic $D(l_{\min},l_{\max})$ to measure the closeness of the CCDF $S(l;l_{\min},l_{\max})$ obtained from the data in the range $l_{\min} \le l \le l_{\max}$ and the theoretical CCDF $P(l;\hat{\eta}(l_{\min},l_{\max}),l_{\min},l_{\max})$ represented by Equation (S2).

$$D(l_{\min},l_{\max}) = \max_{l_{\min}\le l\le l_{\max}} \left|S(l;l_{\min},l_{\max}) - P(l;\hat{\eta}(l_{\min},l_{\max}),l_{\min},l_{\max})\right| \quad \text{(S24)}$$

If $l_{\max} = \hat{l}_{\max}$ is fixed, then $D(l_{\min},\hat{l}_{\max})$ is a function of $l_{\min}$. $l_{\min}$ that minimizes $D(l_{\min},\hat{l}_{\max})$ is numerically chosen from the observed data. That is, $\hat{l}_{\min} = \arg\min_{l_{\min}} D(l_{\min},\hat{l}_{\max})$. From the above equations, $\hat{l}_{\min}$ and $\hat{l}_{\max}$ are obtained. Finally, the exponent $\hat{\eta} = \hat{\eta}(\hat{l}_{\min},\hat{l}_{\max})$ of the TP model that best fits the data in the range $\hat{l}_{\min} \le l \le \hat{l}_{\max}$ is determined using Equation (S3).

## Fitting to EP

In this section, the aim is to determine the minimum value $\hat{l}_{\min}$ of the observed data to be fitted to the EP model and the exponent $\hat{\lambda}$ of the EP model that best fits the data in the range $\hat{l}_{\min} \le l$. The EP in the range $l_{\min} \le l$ is expressed by the following equation:

$$p\left(l;\lambda,l_{\min}\right) = \lambda e^{-\lambda\left(l-l_{\min}\right)} \tag{S25}$$

The CCDF of $p\left(l;l_{\min},\lambda\right)$ is expressed as follows:

$$P\left(l;\lambda,l_{\min}\right) = e^{-\lambda\left(l-l_{\min}\right)} \tag{S26}$$

If the data in the range of $l_{\min} \le l$ are $\left\{l_1, l_2, \cdots, l_m\right\}$, then the log likelihood for these data is expressed as

$$L\left(\lambda;l_{\min}\right) = \sum_{i=1}^{m} \ln p\left(l_i;\lambda,l_{\min}\right) = m\ln\lambda - \lambda\sum_{i=1}^{m}\left(l_i - l_{\min}\right) \tag{S27}$$

The exponent $\hat{\lambda}\left(l_{\min}\right)$ that maximizes $L\left(\lambda;l_{\min}\right)$ is obtained as a solution to $\frac{\partial L\left(\lambda;l_{\min}\right)}{\partial\lambda} = 0$ via the following formula:

$$\hat{\lambda}\left(l_{\min}\right) = m\left(\sum_{i=1}^{m}\left(l_i - l_{\min}\right)\right)^{-1} \tag{S28}$$

The Kolmogorov–Smirnov statistic $D\left(l_{\min}\right)$ is used to measure the closeness of the CCDF $S\left(l;l_{\min}\right)$ obtained from the data in the range $l_{\min} \le l$ and the theoretical CCDF $P\left(l;\hat{\lambda}\left(l_{\min}\right),l_{\min}\right)$ represented by Equation (S6).

$$D\left(l_{\min}\right) = \max_{l_{\min} \le l}\left|S\left(l;l_{\min}\right) - P\left(l;\hat{\lambda}\left(l_{\min}\right),l_{\min}\right)\right| \tag{S29}$$

$\hat{l}_{\min}$ is calculated from $\hat{l}_{\min} = \underset{l_{\min}}{\arg\min}\, D\left(l_{\min}\right)$. The final value is $\hat{\lambda} = \hat{\lambda}\left(\hat{l}_{\min}\right)$.

# Comparison of TP and EP

In this section, the authors describe a method for determining the most suitable distribution model (TP or EP) for the simulation data. Akaike information criterion weights (AICw) are used for comparison [6]. First, the Akaike information criterion (AIC) for data in the range $l_{\min} \leq l \leq l_{\max}$ is defined as follows:

$$
\begin{aligned}
AIC_{TP} &= -2\ln\left(L\left(\hat{\eta}; l_{\min}, l_{\max}\right)\right) + 2 \\
AIC_{EP} &= -2\ln\left(L\left(\hat{\lambda}; l_{\min}\right)\right) + 2
\end{aligned}
\tag{S30}
$$

The AIC difference $\Delta$ is thereafter calculated as follows:

$$
\begin{aligned}
AIC_{\min} &= \min\left(AIC_{TP}, AIC_{EP}\right) \\
\Delta_{TP} &= AIC_{TP} - AIC_{\min} \\
\Delta_{EP} &= AIC_{EP} - AIC_{\min}
\end{aligned}
\tag{S31}
$$

Finally, AICw is calculated as follows:

$$
\begin{aligned}
w_{TP} &= \frac{e^{-\Delta_{TP}/2}}{e^{-\Delta_{TP}/2} + e^{-\Delta_{EP}/2}} \\
w_{EP} &= \frac{e^{-\Delta_{EP}/2}}{e^{-\Delta_{TP}/2} + e^{-\Delta_{EP}/2}}
\end{aligned}
\tag{S32}
$$

First, using the data in the range $\hat{l}_{\min} \leq l \leq \hat{l}_{\max}$ calculated during the fitting of TP, the most appropriate exponents, $\hat{\eta}$ and $\hat{\lambda}$, are determined for each model.

Next, these exponents are used to calculate and compare AICw values. We subsequently changed $\hat{l}_{\min}$ to the one calculated during the fitting of the EP and compared it. If $w_{TP} > w_{EP}$ for both data, TP is considered to fit the simulated data better. However, if $w_{TP} < w_{EP}$ for both data, EP is considered to fit the simulated data better. In the case of discrepancies between the results in both datasets, the following indicators are defined and examined according to reference [2].

$$
\begin{aligned}
D_{adj,TP} &= \frac{\ln N}{\ln n_{TP}} D_{TP} \\
D_{adj,EP} &= \frac{\ln N}{\ln n_{EP}} D_{EP}
\end{aligned}
\tag{S33}
$$

where $D_{\mathrm{TP}}$ and $D_{\mathrm{EP}}$ are the Kolmogorov–Smirnov statistics calculated during the model fitting of TP and EP, respectively. $N$ is the total number of observed data points, and $n_{TP}$ and $n_{EP}$ are the number of observed data points used in each model fitting. In other words, the index considers a model that can fit more observational data. In the case of $D_{adj,TP} < D_{adj,EP}$, TP is considered to fit the simulation data better. Conversely, when $D_{adj,TP} > D_{adj,EP}$, EP is considered to fit the simulation data better.

# REFERENCES

1. Shinohara S, Manome N, Nakajima Y, Gunji YP, Moriyama T, Okamoto H, et al. Power Laws Derived from a Bayesian Decision-Making Model in Non-Stationary Environments. Symmetry. 2021;13(4): 718. doi: 10.3390/sym13040718.
2. Humphries NE, Weimerskirch H, Queiroz N, Southall EJ, Sims DW. Foraging success of biological Lévy flights recorded in situ. Proc Natl Acad Sci U S A. 2012;vol. 109, 19: 7169-7174. doi:10.1073/pnas.1121201109.
3. Jansen VA, Mashanova A, Petrovskii S. Comment on "Lévy walks evolve through interaction between movement and environmental complexity". Science. 2012 Feb 24;335(6071): 918; author reply: 918. doi: 10.1126/science.1215747, PMID: 22362991.
4. White EP, Enquist BJ, Green JL. On estimating the exponent of power-law frequency distributions. Ecology. 2008 Apr;89(4): 905-912. doi: 10.1890/07-1288.1. Erratum in: Ecology. 2008 Oct;89(4): 905–912. DOI: 10.1890/07-1288.1, PMID: 18481513.
5. Clauset A, Shalizi CR, Newman MEJ. Power-Law Distributions in Empirical Data. In SIAM Rev. 2009;51(4): 661–703. DOI: 10.1137/070710111.
6. Edwards AM, Phillips RA, Watkins NW, Freeman MP, Murphy EJ, Afanasyev V, et al. Revisiting Lévy flight search patterns of wandering albatrosses, bumblebees and deer. Nature. 2007 Oct 25;449(7165): 1044-1048. doi: 10.1038/nature06199, PMID: 17960243.